\documentclass[twocolumn,showpacs,preprintnumbers,amsmath,amssymb]{revtex4}
\usepackage{graphicx}% Include figure files
\usepackage{dcolumn}% Align table columns on decimal point
\usepackage{bm}% bold math

\RequirePackage{cmap}
\RequirePackage[cp1251]{inputenc}
\RequirePackage[TS1,T2A]{fontenc}
\newcommand{\frat}[2]{\frac{\textstyle #1}{\textstyle #2}}
\newcommand{\vf}[1]{\mbox{\boldmath $#1$}}

\begin{document}

\title{Quantum liquids resulted from the models with four-fermion
interaction}

\author{S. V. Molodtsov}
 \altaffiliation[Also at ]{
Institute of Theoretical and Experimental Physics, Moscow, RUSSIA}
\affiliation{%
Joint Institute for Nuclear Research, Dubna,
Moscow region, RUSSIA%\textbackslash\textbackslash
}%
\author{G. M. Zinovjev}
\affiliation{
Bogolyubov Institute for Theoretical Physics,
National Academy of Sciences of Ukraine, Kiev, UKRAINE
}%

\date{\today}% It is always \today, today,

\begin{abstract}
A (nearly) perfect liquid discovered in the experements with
ultrarelativistic heavy ion collisions is investigated by studying the
quark ensembles with four-fermion interection as a fundamental theoretical
approach. The comparative analysis of several quantum liquid models is
performed and it results in the conclusion that the presence of gas---liquid
phase transition is their characteristic feature. The problem of instability
of small quark number droplets is discussed and argued it is rooted in the
chiral soliton formation. An existence of mixed phase of the vacuum and
baryon
matter is proposed as a possible reason of the latter stability.

\end{abstract}

\maketitle

Huge amount of data on relativistic heavy ion collisions obtained recently
(perceptibly before the LHC began operating) in various experiments (first of
all, at RHIC), were well understood and described in terms of concepts based on
the equations of relativistic hydrodynamics~\cite{hydro}. In particular, nearly
ideal hydrodynamics, supplemented as needed by a variety of hadronic cascade models
so as to correctly take into account a hadronic stage of the
collision~\cite{teaney}, quite successfully predicted an appearance of the radial and elliptic flows,
their dependence on the mass, centrality, beam energy, and transverse momentum
(though restricted in the magnitude), clearly indicating at the same time that the
expanding liquid exhibits sufficiently specific transport properties. It is very much
close to the ideal one, since the ratio $\eta/s$ of its shear viscosity coefficient
$\eta$ to the density entropy $s$ turned out to be a small quantity.

At this point, it should be mentioned that the exploitation of such
hydrodynamic notions dates back to the early fifties of the last century when L. D. Landau
had developed the model of multiple particle production in collisions of hadrons
and nuclei guided by the hydrodynamics in describing the evolution of nuclear
matter that occurs right upon squeezing the latter at the collision point~\cite{landau}.
Conceptually, this breakthrough idea had not been particularly successful in
applications then because the nuclear matter had turned out to be a not very
“suitable”\ liquid (as it was considered at the time), as the mean free path of nucleons
in a produced system was fully comparable with the size of the latter.

A new generation of experiments carried out at much higher energies (reached
at the LHC) quite remarkably confirmed predictions obtained by applying the past
hydrodynamic ideas, rendering some of the latter, for instance, an observation of higher
harmonics of flow induced by the fluctuations of original geometry, or a jet quenching effect
initiated by heavy and light quarks, to be not only reliable experimental
data~\cite{jetflow}, but also observations that bear a profound heuristic meaning.

The physics of ultrarelativistic heavy ion collisions needs to be described,
at least, at the initial stage, in the language of quantum chromodynamics (QCD) for a
strongly interacting system that is in a state far from equilibrium. At the same time,
the data obtained by all three LHC collaborations, while being successfully described
in terms of hydrodynamics, suggest a very fast thermalization, i.e., sufficient degree of
local equilibrium or, rather, isotropization, since the equations of hydrodynamics
do not include the temperature of produced matter with explicit collective properties, whose
theoretical explanation at the macroscopic level is still far from to be clear. In recent
years, there have appeared several scenarios of what could be the dynamics of a system
transiting from the initial collision state to that when it becomes (almost)
equilibrium~\cite {blaizot}. However, this problem is not discussed in the present paper. Instead, we
focus on another aspect of the problem, namely, the smallness of the ratio $\eta /s$ which
corresponds to the presence of the strong interaction in a produced system or, in other
words, small mean free path of its constituents, and try to understand the very nature of such
interactions in a system, whose dynamics is governed by the coupling constant, which is
likely not too large (at the LHC energies the running coupling constant in QCD as
$\alpha_s\sim 0.3$---$0.4$) avoiding AdS/CFT duality (holographic QCD) arguments very popular at the
moment~\cite{herzog}.

Recently, such multiparticle (fermionic) systems are being intensively
studied, particularly after they have successfully been realized in experiments as ultracold gas of
fermionic atoms \cite{coldatom}. This (unitary) Fermi-gas is a dilute system with short-range
interaction, in which the $s$-wave scattering among fermions saturates a unitary limit for
the cross section. Such a system is naturally characterized by the absence of any internal scale
(conformality) and does not depend on the details of interaction. On the other hand, the
interaction in such a system needs to be described non-perturbatively, since no small parameter
exists in the problem. An ideal liquid observed in heavy ion experiments is exactly another
remarkable example of such a strongly correlated fermionic system. An assumption, that
there exists the lower bound for the ratio $\eta/s$ of such fermionic systems formulated
in~\cite{KSS}, has triggered even greater interest in their study after it has been shown that
$\eta/s$ for the systems produced in heavy ion collisions and ultracold atomic gases turns out
to be very small and close to each other in magnitude. It is interesting that the same value
of that ratio is also predicted for low-energy electrons in graphene
monolayers~\cite{graphene}. The nature of these phenomena is, however, unclear which
is seen from the behavior of, say,
volume viscosity that for the quark—gluon systems turns out to be nonzero and can under
certain circumstances (nearby phase transitions) be a significant source of dissipation, whereas
for the unitary Fermi-gases it vanishes, just as a consequence of the scale invariance.

The four-fermion (QCD-like) field theories still remain a most reliable
source of quantitative information in the studies of the transport properties of strongly correlated
systems and their thermodynamics, in particular, a chiral phase transition between
massive hadrons and almost massless quarks. It is a thermodynamics that provides us with some
general framework which lets one to understand how the properties of macroscopic matter and, in
particular, its collective behavior, emerge from the laws that govern microscopic dynamics.
The results of this work allows us to suppose with a sufficient, in our view, level of
argumentation that the picture based on a complex collective behavior of quarks (antiquarks,
gluons), which is expressed in the presence of vacuum condensates even under normal conditions,
can be set by the nontrivial thermodynamic properties of vacuum, which eventually determine
the observable properties of strongly interacting matter. In our opinion, this possibility
was not sufficiently widely discussed and, even more so, used already at the initial
stage of studies of the quark—gluon matter due to purely accidental circumstances. (As recent
discussions of one of us (G.M.Z.) with E. Shuryak that have taken place during "Quark Matter
2012"\ have shown, similar thoughts are fully shared by him and, presumably, have occurred to
him a bit earlier (see, for example, \cite{ESTS}.)

%%%%%%%%%%%%%%%%%%%%%%%%%%%%%%%%%%%%%%%%%%%%%%%%%
\begin{figure}%[!tbh]
\includegraphics[width=0.3\textwidth]{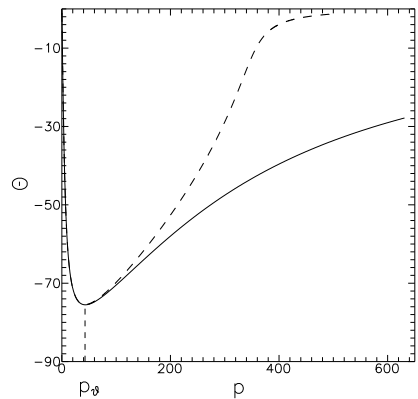}
\caption{The most stable equilibrium angles $\theta$ (in degrees) as
function of momentum $p$ in MeV. The solid line for NJL model, dashed one
corresponds to the KKB model.
}
%1
\label{f1}
\end{figure}
%%%%%%%%%%%%%%%%%%%%%%%%%%%%%%%%%%%%%%%%%%%%%%%%%
\section{Thermodynamics of the ensemble}
In the present work we consider some aspects of thermodynamical description
of the quark ensemble with four-fermion interaction (generated, as it is believed, by
strong stochastic gluon field) Hamiltonian density
\begin{equation}
%1
\label{1}
{\cal H}=-\bar q~(i{\vf \gamma}{\vf \nabla}+m)~q-j^a_\mu \int d{\vf y}~
\langle A^{a}_\mu A'^{b}_\nu\rangle~j'^b_\nu~,
\end{equation}
where $j^a_\mu=\bar qt^a\gamma_\mu q$ is the quark current, with
corresponding quark operators $q$, $\bar q$, taken in spatial point ${\vf x}$ (the variables with
prime corresponds to the ${\vf y}$ point), $m$ is the current quark mass, $t^a=\lambda^a/2$ is
the color gauge group $SU(N_c)$ generators, $\mu,\nu=0,1,2,3$. We take the gluon field
correlator $\langle A^{a}_\mu A'^{b}_\nu\rangle$ in a simple form of color singlet, with
contact (in time) interaction (without retarding){\footnote{Generally speaking, in such a
correlation function the terms spanned on the vector of relative distance are allowed, but for
simplicity we ignore them.}}
\begin{equation}
%2
\label{cor}
\langle A^{a}_\mu A'^{b}_\nu\rangle=G~\delta^{ab}~\delta_{\mu\nu}~
F({\vf x}-{\vf y})~,
\end{equation}
(we do not include corresponding delta-function on time in this formula).
This simple correlation function is a fragment of corresponding ordered exponent and
besides the four-fermion interaction accompanied infinite number of multi-fermion
vertices arises. But for our purposes here it would be quite enough restrict ourself with this simple
form. The mentioned above effective interactions appear in natural way by the coarse-grained
description of the system with handling the corresponding averaging procedure, and having
in mind that vacuum gluon field changed stochastically (for example, in the form of
instanton liquid, see~\cite{MZ}). But this elaboration of effective Hamiltonian resulting from
the first principles of quantum chromodynamics (QCD) will be unessential for us, as it
will be demonstrated below. The choice of correlation function in the simplest form
with instantaneous interaction does not generate any problem at transforming final results from
the Minkovski space to the Euclidean one and the formfactor $F({\vf x})$ is interpreted in
a simple way as  an interaction 'potential' of point-like particles. The correlation function
itself looks, formally, like a gauge non-invariant object{\footnote{It is obvious that we
are telling about an approximate calculation of corresponding generating functional for some
specific conditions with the restricted area of applicability that does not imply the calculation
of functional derivatives of arbitrary order.}}. Nevertheless, there exists an effective
way to significantly compensate for this shortcoming, if all similar 'potentials' are looked
through, in some sense, (to be elucidated below). For example, this set would be quite perceptible,
if it becomes  possible to confront two limits opposite in physics, for example, started
from the formfactor with a delta-like function in the coordinate space (the Nambu--Jona--Lasinio
(NJL) model~\cite{njl}, the correlation length is finite in this case) and extended to
the formfactor of a delta-like function in the momentum space (clearly, the correlation length
tends to infinity in this case) analogous to that is well-known in condensed matter physics as the
Keldysh model (KKB) \cite{kldsh}.
%%%%%%%%%%%%%%%%%%%%%%%%%%%%%%%%%%%%%%%%%%%%%%%%%
\begin{figure}%[!tbh]
\includegraphics[width=0.3\textwidth]{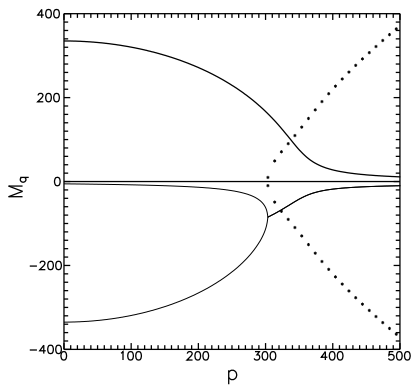}
\caption{Three branches of solutions for dynamical quark mass (in MeV) for
the KKB model as a function of momentum (MeV). The imaginary parts of the
solutions are shown by dots.}
%2
\label{f2}
\end{figure}
%%%%%%%%%%%%%%%%%%%%%%%%%%%%%%%%%%%%%%%%
It is worth to remark here that we will need only one of
its properties, although {\footnote{In the KKB model the fermion behavior is considered in
the stochastic random field with infinite correlation length (the NJL model corresponds to
the 'white noise' with zero correlation length). In this case one is lucky enough to be able to
'sum up' an entire diverging series and, therefore, to demonstrate that fermions are in
general not on mass shell.}} of exceptional importance, which is related to the fact that, due to
the special formfactor behavior, all the momentum integrations in the problem get
factorized and effectively the problem becomes one-dimensional (then only integration over
energy are in play). From this point of view, other models with an arbitrary formfactor (including
the NJL model) could be represented as a superposition of elementary blocks obtained by
using the KKB model. The utmost distributions mentioned above can be considered as a limiting case
for the corresponding Gaussian correlators in the coordinate and momentum spaces,
which, of course, look more realistic. The coupling constant scale $G$, that will turn out to
be interesting for applications can be tuned by using corresponding PDG meson observables.
Comparing the results obtained (by continuity arguments) one can make some conclusions about
behavior of the system with practically any interaction potential.

We consider necessary to comment briefly on a case with a linear potential,
which was always giving hope to discover an unusual feature in quark behavior thereby shedding
some light on the nature of confinement. Meanwhile, at present however, it appears that such a
singular 'potential' is even superfluous for our purposes, since the properties, we
are interested in, are already revealed in the KKB model, which, in a sense, is like half way
from the NJL model to that with a linear potential. Secondly, the quasiparticles in the model
with a linearly increasing potential can not basically be distinguished from those in, for
example, the NJL model, provided an integrable infrared singularity in the former is
eliminated. As a result the same massive objects appear without the anomalies in
the energy spectrum. Additionally, the analysis shows that the multi-fermion contributions present
in the problem in a general case can be reduced to the four-fermion interaction in an
acceptable way by inserting the respective vacuum expectation values. In other words, even the
Hamiltonian of the form (\ref{1}) seems to capture the essential features of quark interactions.
%%%%%%%%%%%%%%%%%%%%%%%%%%%%%%%%%%%%%%%%%%%%%%%%%
\begin{figure}%[!tbh]
\includegraphics[width=0.3\textwidth]{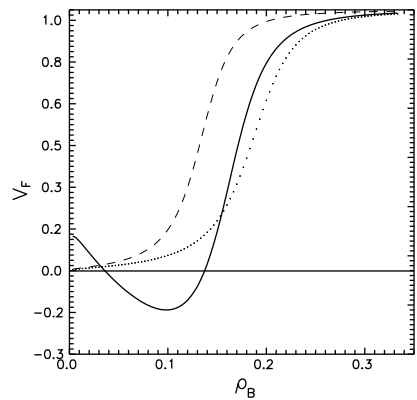}
\caption{The group velocity of quasiparticles $v_{\mbox{\tiny{F}}}$ on
the Fermi surface. The solid line describes the NJL model, the dashed one
corresponds to the KKB model, the dots show the data for the KKB model tuned to the
$\pi$-meson energy Figs.\ref{f4}--\ref{f8}).}
%3
\label{f3}
\end{figure}
%%%%%%%%%%%%%%%%%%%%%%%%%%%%%%%%%%%%%%%%%

It is believed that at sufficiently large interaction the ground state of the
system transforms from a trivial vacuum $|0\rangle$ (the vacuum of free Hamiltonian) to the
mixed state (with quark--anti-quark pairs with the opposite momenta and vacuum quantum numbers)
which is presented as the Bogolyubov trial function (in that way some separate
reference frame is introduced and a chiral phase becomes fixed)
$$
|\sigma\rangle={\cal{T}}|0\rangle,~
{\cal{T}}=\prod\limits_{ p,s}\exp[\varphi_p~(a^+_{ p,s}b^+_{- p,s}+
a_{ p,s}b_{-p,s})].$$
Here $a^+$, $a$ и $b^+$, $b$ are the quarks creation and annihilation
operators, $a|0\rangle=0$, $b|0\rangle=0$. The dressing transformation ${\cal{T}}$ transmutes the quark
operators to the creation and annihilation operators of quasiparticles
$A={\cal{T}}~a~{\cal{T}}^\dagger$,
$B^+={\cal{T}}~b^+{\cal{T}}^\dagger$.

The termodynamic properties of a quark ensemble are known to be determined by
solving the following problem. It is required to find such a statistical operator
\begin{equation}
%3
\label{dm}
\xi=\frat{e^{-\beta ~\hat H_{{\mbox{\scriptsize{app}}}}}}{Z_0}~,
~~Z_0=\mbox{Tr}~\{e^{-\beta ~\hat H_{{\mbox{\scriptsize{app}}}}}\}~~,
\end{equation}
that at fixed mean charge
\begin{equation}
%4
\label{ntot}
\overline{Q}_0=\mbox{Tr} \{\xi ~Q_0\}=
V~\gamma~\int  d \widetilde{\vf p}~(n-\bar n)~,
\end{equation}
$d \widetilde {\bf p}=d {\bf p}/(2\pi)^3$,
($Q_0=\bar q \gamma^0 q$), and fixed mean entropy
\begin{eqnarray}
%5
\label{stot}
\overline{S}&=&-\mbox{Tr} \{\xi~ S\}=\\
&-&V~\gamma~\int d \widetilde {\vf p}~
[n\ln n+(1-n)\ln (1-n)+\nonumber\\
&+&\bar n\ln \bar n+(1-\bar n)\ln (1- n)],\nonumber
\end{eqnarray}
($S=-\ln \xi$), provides a minimal value of mean energy of the quark ensemble
$$E=\mbox{Tr} \{\xi~H\}~,$$
($H=\int d{\vf x}~ {\cal H}$). In other words, we are interested in the
minimum of the
following functional
\begin{equation}
%6
\label{fun}
\Omega=E-\mu~\overline{Q}_0 -T~\overline{S}~,
\end{equation}
where $\mu$ and $T$ denote the Lagrangian multipliers for the chemical
potential of the quark/baryon charge (which is usually taken to be three times larger than the
baryon one in phenomenological considerations) and the temperature ($\beta=T^{-
1}$), respectively. $V$ is the volume the sytem is enclosed in,
$\gamma=2 N_c$ (in the case of several quark flavors $\gamma=2 N_c N_f$, where $N_f$ is the
flavor number), $n=\mbox{Tr} \{\xi A^+ A\}$, $\bar n=\mbox{Tr} \{\xi B^+ B\}$ are the
components of the corresponding density matrix.
%%%%%%%%%%%%%%%%%%%%%%%%%%%%%%%%%%%%%%%%%%%%%%%%%
\begin{figure}%[!tbh]
\includegraphics[width=0.3\textwidth]{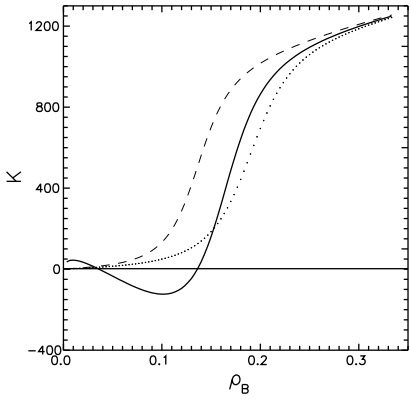}
\caption{The compression module $K$ in MeV.
}
%4
\label{f4}
\end{figure}
%%%%%%%%%%%%%%%%%%%%%%%%%%%%%%%%%%%%%%%%%

We restrict ourselves by considering the Bogolyubov--Hartree--Fock
approximation in which the statistical operator is constructed on the basis of approximating effective
Hamiltonian $H_{{\mbox{\scriptsize{app}}}}$, quadratic in creation and annihilation
operators for quasiparticles acting in the corresponding Fock space with a vacuum state
$|\sigma\rangle$. The average specific energy per quark $w=E/(V\gamma)$ results in \cite{MZ2}
\begin{eqnarray}
%7
\label{w}
w&=&\int d \widetilde{\vf p}~p_0-\int d \widetilde{\vf p}~
(1-n-\bar n)~p_0~\cos\theta-
\nonumber\\[-.2cm]
\\ [-.25cm]
&-&\frat12~\int d \widetilde{\vf p}~(1-n-\bar n)
\sin \left(\theta-\theta_m\right)~M({\vf p})~,\nonumber
\end{eqnarray}
where
$$
M({\vf p})=2G\int d \widetilde{\vf q}~(1-n'-\bar n')~
\sin \left(\theta'-\theta'_m\right)~F({\vf p}+{\vf q})~,$$
$\theta=2\varphi$, $p_0=({\vf p}^2+m^2)^{1/2}$, the primed variables,
hereinafter correspond to the integration over momentum ${\vf q}$. The auxiliary angle $\theta_m$ is
determined from the relation $\sin \theta_m=m/p_0$. The first term in Eq.(\ref{w}) is introduced
in view of normalizing in such a way to have the zero energy of ground state when an
interaction is switched off. This constant is unessential for the following consideration
and may be omitted, however it should be kept in mind that it will appear as a regularizer in
singular expressions further down the text.

The most stable extremals of the functional (\ref{w}) are presented for
comparison with the solid line for the NJL model and dashed one for the KKB model under normal
conditions ($T=0$, $\mu=0$) in Fig.\ref{f1}. For the delta-like potential in coordinate space
(the NJL model) the expression (\ref{w}) diverges and to obtain the reasonable results the upper
limit cutoff in the momentum integration $\Lambda$ is introduced being one of the tuning
model parameters along with the coupling constant $G$ and current quark mass $m$. Below, we use one
of the standard sets of the parameters for the NJL model \cite{5}: $\Lambda=631$ MeV,
$G\Lambda^2/(2\pi^2)\approx 1.3$, $m=5.5$ MeV, whereas the KKB model
parameters are chosen in such a way that for the same quark current masses the dynamical quark ones in
both NJL and KKB models coincide at vanishing quark momentum. The momentum $p_\vartheta$
(parameter) corresponds to the maximal attraction between quark and anti-quark. The value of this
parameter inversed determines a characteristic size of quasiparticle.
%%%%%%%%%%%%%%%%%%%%%%%%%%%%%%%%%%%%%%%%%%%%%%%%%
\begin{figure}%[!tbh]
\includegraphics[width=0.3\textwidth]{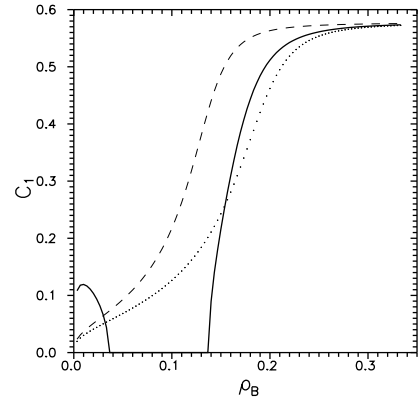}
\caption{The first sound velocity $C_1$.
}
%5
\label{f5}
\end{figure}
%%%%%%%%%%%%%%%%%%%%%%%%%%%%%%%%%%%%%%%%%
It is of order of $p_\vartheta \sim(mM_q)^{1/2}$, where $M_q$ is a
characteristic quark dynamical mass for the models considered, i.e. the quasiparticle size is
comparable with the size of $\pi$-meson (Goldstone particle). It is a remarkable fact that the
quasiparticle, as it is seen from Fig. \ref{f1}, does not depend noticeably on the formfactor
profile or, in other words, on the scale, but rather depends on the coupling constant. Using the
properties of extremals the functional expression (\ref{w}) can be transformed to the form
(see \cite{MZ2})
\begin{eqnarray}
%8
\label{w2}
w&=&\int d \widetilde{\vf p}~p_0-\int d \widetilde{\vf p}~(1-n-\bar n)~P_0+
\nonumber\\[-.2cm]
\\ [-.25cm]
&+&\frat{1}{4G}~\int d \widetilde{\vf p}d \widetilde{\vf q}~~F({\vf p}+{\vf q})
~\widetilde M({\vf p})
\widetilde M({\vf q})~,\nonumber
\end{eqnarray}
where $P_0=[{\vf p}^2+M_q^2({\vf p})]^{1/2}$ is the energy of quark
quasiparticle with the dynamical quark mass
\begin{equation}
%9
\label{9}
M_q({\vf p})=m+M({\vf p})=m+\int d \widetilde{\vf q}~
F({\vf p}+{\vf q})~\widetilde M({\vf q})~.
\end{equation}
Below we omit often the arguments of corresponding functions for the mass and
quasiparticle energy. Varying the functional (\ref{w2}) with respect to the density of
induced quasiparticle mass $\widetilde M$ (in such a form it is convenient to take variational
derivatives{\footnote{If one decides to take the dynamical quark mass $M_q$, as a basic
variable, then it is seen from Eq.(\ref{9}) that formulating an inverse transformation from
$M_q$ to $\widetilde M$ suitable for handling is difficult.}}) we obtain the equation
for dynamical quark mass as
\begin{equation}
%10
\label{10}
M_q({\vf p})=m+2G\int d \widetilde{\vf q}~
(1-n'-\bar n')~\frat{M'_q}{P'_0}~F({\vf p}+{\vf q}),
\end{equation}
which corresponds exactly to the mean field approximation. In particular,
under normal conditions ($T=0$, $\mu=0$) the dynamical quark mass in the NJL model is
$M_q\sim 340$ MeV, whereas the dynamical quark mass of the KKB model is determined by the
following equation
\begin{equation}
%11
\label{mkeld}
M({\vf p})=2 G~\frat{M_q({\vf p})}{P_0}~.
\end{equation}
In practice, it is convenient to use an inverse function $p(M_q)$. Then in
the chiral limit $M_q=(4G^2-{\vf p}^2)^{1/2}$, at $|{\vf p}|<2G$, and $M_q=0$ when
$|{\vf p}|>2G$. In this case the quark states with momenta $|{\vf p}|<2G$ are degenerate in energy
$P_0=2G$. Fig.\ref{f2} demonstrates three branches of the equation (\ref{mkeld}) solutions for
dynamical quark mass. The dots show the imaginary part of solutions which are generated at the
point where two real solution branches are getting merged.
%%%%%%%%%%%%%%%%%%%%%%%%%%%%%%%%%%%%%%%%%%%%%%%%%
\begin{figure}%[!tbh]
\includegraphics[width=0.3\textwidth]{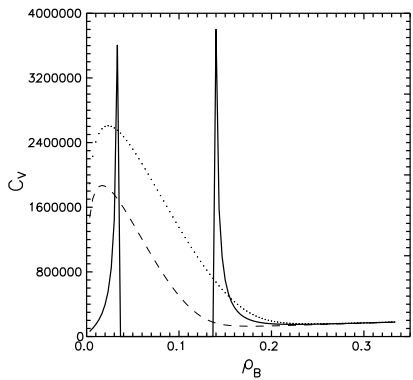}
\caption{The slope factor at low temperatures $\frat13 \pi^2
N_{\mbox{\tiny{F}}}$ in the thermal conductivity expression at constant
volume $C_{\mbox{\scriptsize V}}=\frat13 \pi^2 N_{\mbox{\tiny{F}}} T$.
}
%6
\label{f6}
\end{figure}
%%%%%%%%%%%%%%%%%%%%%%%%%%%%%%%%%%%%%%%%%

%%%%%%%%%%%%%%%%%%%%%%%%%%%%%%%%%%%%%%%%%%%%%%%
%%%%%%%%%%%%%%%%%%%%%%%%%%%%%%%%%   2
%%%%%%%%%%%%%%%%%%%%%%%%%%%%%%%%%%%%%%%%%%%%%%%
\section{Mean energy as a functional of quantum liquid theory}
The goal that we pursued while passing from the expression for specific
energy (\ref{w}) to Eq. (\ref{w2}) was to derive such a form that would easily be recognized as an
energy functional of the Landau Fermi-liquid theory \cite{lfl}. Some aspects of this theory are
interesting and useful to be applied for comparing the results obtained in the NJL and KKB
models. We will also discuss the first order phase transition which is apparently typical for
interacting fermions (relativistic Fermi-liquid).

Thus, the second term in (\ref{w2}) describes the contributions coming from
quark and antiquark quasiparticles with occupation numbers $n$ and $\bar n$ respectively. The
unity in the expression $1-n'-\bar n'$ corresponds to the vacuum fluctuations of quarks
and antiquarks. The last term in (\ref{w2}) is
due to the interaction of quasiparticles. The presence of contributions coming from
antiparticles and the relativistic form of dynamics are those features which distinguish quark
ensembles we study from the Fermi-liquids considered in condensed matter physics. The first
variation of the functional (\ref{w2}) with respect to the particle (antiparticle) density
leads (as it should be) to the energy of quasiparticle:
\begin{equation}
%12
\label{1var}
\frat{\delta w}{\delta n}=P_0~.
\end{equation}

Consider, first, the situation of zero temperature and discuss some aspects
of filling up the Fermi sphere by quarks. Let us assume that the momentum distribution of
quarks (antiquarks) is determined by the following expressions taken at the $\beta \to 0$ limit
\begin{equation}
%13
\label{newden}
n=\left [e^{\beta(P_0-\mu)}+1\right]^{-1},
~\bar n=\left[e^{\beta(P_0+\mu)}+1\right]^{-1},
\end{equation}
that is by the Fermi step function: $n=1$, at $P_0\leq \mu$ and $n=0$ when
$P_0>\mu$. It is clear that for antiquarks $\bar n=0$. The quark density is determined by
using the Fermi momentum:
\begin{equation}
%14
\label{den}
\rho=\frat{\gamma P_{\mbox{\tiny{F}}}^3}{6\pi^2}~,~~\rho=\frat{Q_0}{V}~,
\end{equation}
with the quark chemical potential that coincides with the quasiparticle
energy on the Fermi surface, as it follows from the relation (\ref{1var}), i.e.
\begin{equation}
%15
\label{chem}
\mu=[P^2_{\mbox{\tiny{F}}}+M_q^2(P_{\mbox{\tiny{F}}})]^{1/2}~.
\end{equation}
%%%%%%%%%%%%%%%%%%%%%%%%%%%%%%%%%%%%%%%%%%%%%%%%%
\begin{figure}%[!tbh]
\includegraphics[width=0.3\textwidth]{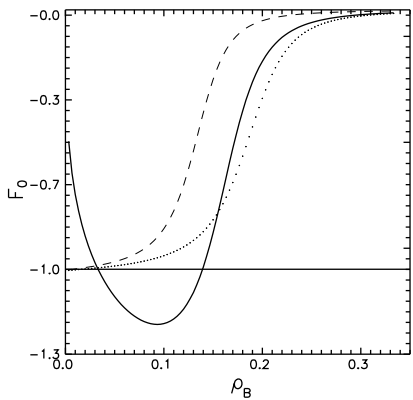}
\caption{The factor $F_0$ of the Landau Fermi-liquid theory.
}
%7
\label{f7}
\end{figure}
%%%%%%%%%%%%%%%%%%%%%%%%%%%%%%%%%%%%%%%%%

The group velocity of quasiparticles on the Fermi surface
$v_f=\partial P_0/\partial {\bf p}|_{|{\bf p}|=P_{\mbox{\tiny{F}}}}$ is shown
in Fig.\ref{f3} as a function of baryon (quark) density (by definition, the baryon density is
three times smaller than the quark one $\rho_{{\mbox{\scriptsize B}}}=\rho/3$). A solid
line describes the NJL model, while a dashed one corresponds to the KKB model. There are points
for comparison that show a version for the KKB model when the parameters are tuned in such a
way that the $\pi$-meson masses coincide in the NJL and KKB models (similar notation is
used below in Figs. \ref{f4}---\ref{f8}). Tending the group velocity to unity in the region
of normal nuclear densities corresponds to the chiral symmetry restoration when an induced
quark mass tends to zero. The group velocity turns to zero for quarks with momenta $|{\bf p}|<2G$
in the chiral limit in the KKB model. The negative group velocities in the NJL model
correspond to the regions of instability (see below). The points in which the group velocity
vanishes give rise to the peaks in the density of states on the Fermi surface $N_{\mbox{\tiny{F}}}$,
\begin{eqnarray}
%28
\label{Nsur}
&&\hspace{-0.5cm}N_{\mbox{\tiny{F}}}=\gamma \int d \widetilde{\bf p}~
\delta (P_0-\mu)=
\frat{\gamma}{2\pi^2}P_{\mbox{\tiny{F}}} P^0_{\mbox{\tiny{F}}}~
\left(1+F_0\right)^{-1}\!\!\!,\\
&&\hspace{-0.5cm}F_0=\frat{M_q}{P_{\mbox{\tiny{F}}}}\frat{dM_q}{dP_{\mbox{\tiny{F}}}
}~,\nonumber
\end{eqnarray}
where $P^0_{\mbox{\tiny{F}}}=\left.P_0\right|_{{\mbox{\small |{\bf p}|= P}}_{\mbox{\tiny{F}}}}$,
$N_{\mbox{\tiny{F}}}=d\rho/d\mu$. For more detail on how to determine the
parameter $F_0$, see below. The interaction term in the functional (\ref{w2}) vanishes in an ideal
gas and causes the derivative of quark dynamical mass in the Fermi momentum to turn to zero:
$d M_q/d P_{\mbox{\tiny{F}}}=0$. Let us define the density of states of an
ideal gas as
$$\widetilde N_{\mbox{\tiny{F}}}=\gamma/(2\pi^2)P_{\mbox{\tiny{F}}}
P^0_{\mbox{\tiny{F}}}~,$$
then the relation (\ref{Nsur}) can be written in the form:
$$N_{\mbox{\tiny{F}}}=\widetilde N_{\mbox{\tiny{F}}}\left(1+F_0\right)^{-1}~.
$$

Another important characteristic is a compression coefficient
\begin{equation}
%29
\label{k}
K=9 \rho~ \frat{d \mu}{d \rho}=3~\frat{P^2_{\mbox{\tiny{F}}}}{\mu}
\biggl(1+F_0\biggr)~.
\end{equation}
Fig. \ref{f4} demonstrates the data for the NJL and KKB models. They are
consistent with the specific values obtained for nuclear medium. One can also conclude that,
in principle, these models admit a wide variety of equations of state including
sufficiently restrictive ones. The negative values of the compression coefficient are not allowed and signal
the region of instability. The first sound velocity which is determined by the relation
\begin{equation}
%30
\label{c1}
C_1^2=\frat{K}{9~\mu}=\frat{v^2_{\mbox{\tiny{F}}}}{3}
\biggl(1+F_0\biggr)~,
\end{equation}
is shown in Fig. \ref{f5}. When baryon densities are somewhat higher than the
density of normal nuclear matter, the sound velocity tends to its asymptotic value
$C_1=1/\sqrt{3}$ which is a natural manifestation of the chiral symmetry restoration. If the sound
velocity of an ideal Fermi-gas $\widetilde C_1^2= v^2_{\mbox{\tiny{F}}}/3$ is introduced in a way
similar to the $\widetilde N_{\mbox{\tiny{F}}}$ definition, then the expressions (\ref{Nsur}), (\ref{c1})
can be endowed with the form whose physical meaning is an equality of flow coming through the
Fermi sphere of quasiparticles of (imaginary) ideal Fermi-gas and interacting Fermi-liquid
(that is, there basically is a relativistic analogue of the Luttinger theorem \cite{lat})
\begin{equation}
%31
\label{lt}
N_{\mbox{\tiny{F}}}~ C^2_1=\widetilde N_{\mbox{\tiny{F}}}~ \widetilde C^2_1~.
\end{equation}
%%%%%%%%%%%%%%%%%%%%%%%%%%%%%%%%%%%%%%%%%%%%%%%%%
\begin{figure}%[!tbh]
\includegraphics[width=0.3\textwidth]{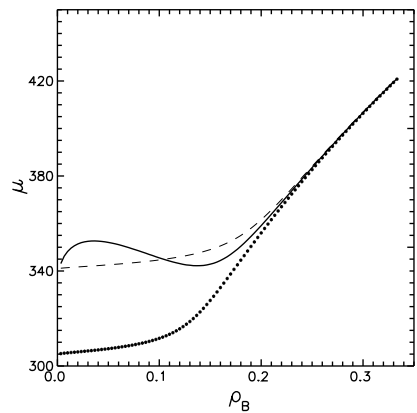}
\caption{Chemical potential in MeV.
}
%8
\label{f8}
\end{figure}
%%%%%%%%%%%%%%%%%%%%%%%%%%%%%%%%%%%%%%%%%

The thermal conductivity at a constant volume and a low temperature is given
by the expression
\begin{equation}
%32
\label{cv}
C_{\mbox{\scriptsize V}}=\frat13~\pi^2 N_{\mbox{\tiny{F}}}~ T~.
\end{equation}
Fig. \ref{f6} shows the slope (the factor $\frat13 \pi^2 N_{\mbox{\tiny{F}}}$
in Eq.(\ref{cv}), $N_{\mbox{\tiny{F}}}=d \rho/d \mu$), as a function of baryon/quark density
that demonstrates how informative it could be to measure the slope of a curve corresponding to
the thermal conductivity. Yet another important characteristic of Fermi-liquid is defined
by the second variational derivative, which for the functional (\ref{w2}) develops only a
scalar component
\begin{equation}
%33
\label{v2}
f_0=\frat{\delta^2 w}{\delta n^2}=\frat{M_q}{P_0}\frat{\delta M_q}{\delta n}~.
\end{equation}
For the Fermi-liquid at zero temperature, in particular, we have
$$f_0=\frat{2\pi^2}{\gamma P_{\mbox{\tiny{F}}}~P^0_{\mbox{\tiny{F}}}}
\frat{M_q}{P_{\mbox{\tiny{F}}}}\frat{d M_q}{d P_{\mbox{\tiny{F}}}}~.
$$
For example, in the NJL model
\begin{eqnarray}
&&\frat{M_q}{P_{\mbox{\tiny{F}}}}\frat{d M_q}{d P_{\mbox{\tiny{F}}}}=
-\frat{P_{\mbox{\tiny{F}}}}{P^0_{\mbox{\tiny{F}}}}\frat{1}{I+\pi^2 m/(G
M_q^3)}~,\nonumber\\
&&
I=\ln \frat{\Lambda+P^0_{\tiny{\Lambda}}}{P_{\mbox{\tiny{F}}}+
P^0_{\mbox{\tiny{F}}}}-
\frat{\Lambda}{P^0_{\tiny \Lambda}}+\frat{P_{\mbox{\tiny{F}}}}
{P^0_{\mbox{\tiny{F}}}}~.\nonumber
\end{eqnarray}
where $P^0_{\scriptsize \Lambda}=\left.P_0\right|_{{\small |{\bf p}|= \Lambda}}$. In the KKB model
$$\frat{M_q}{P_{\mbox{\tiny{F}}}}\frat{d M_q}{d P_{\mbox{\tiny{F}}}}=
-\frat{M~M_q^2}{M_q^3+m P_{\mbox{\tiny{F}}}^3}~.$$
In particular, in the chiral limit (when $m=0$) we have
$(M_q/P_{\mbox{\tiny{F}}})(d M_q/d P_{\mbox{\tiny{F}}})=-1$. The collective
oscillation modes of the Fermi-liquid, the so-called zero sound (the collisionless mode), are
found by using the parameter
$$F_0=\widetilde N_{\mbox{\tiny{F}}}~f_0=
\frat{M_q}{P_{\mbox{\tiny{F}}}}\frat{d M_q}{d P_{\mbox{\tiny{F}}}}~,
$$
which is shown in Fig. \ref{f7}. In particular, in the KKB model
$$F_0= -\frat{M M_q^2}{M M_q^2+(P^0_{\mbox{\tiny{F}}})^2 m}\geq -1~.
$$
The zero sound oscillations are known to be determined by the solutions
to the dispersion equation with a frequency parameter $s$ (for details
concerning this notation see the section devoted to the polarization operator) of the form:
\begin{equation}
%34
\label{dzs}
F_0=\frat{s}{2}~\log\frat{s+1}{s-1}-1~.
\end{equation}
When there is a repulsion in a system and the factor is positive $F_0>0$, the
solutions to the dispersion equation $s=\lambda+i\eta$ describe continuous oscillations
($\eta=0$). In the case of weak attraction, when $-1<F_0<0$, the damped oscillations of zero sound
are possible with a purely imaginary frequency ($\lambda=0$) which is given by the solutions to
the following equation:
$$F_0+1=\eta~\arctan(1/\eta)~.$$
When the strong attraction is available and $F_0<-1$, the solutions reside on
a second sheet of the complex plane $s$ and describe the damped oscillations which are found
from the solution to the equation
$$F_0+1=\eta~[-\pi+\arctan(1/\eta)]~.$$
%%%%%%%%%%%%%%%%%%%%%%%%%%%%%%%%%%%%%%%%%%%%%%%%%
\begin{figure}%[!tbh]
\includegraphics[width=0.3\textwidth]{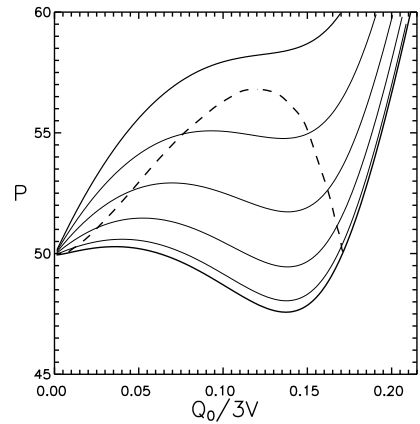}
\caption{The ensemble pressure $P$ (MeV/fm$^3$) is shown as a function of
charge density ${\cal Q}_0$ at temperatures $T=0$ MeV, \dots , $T=50$ MeV
with spacing $T=10$ MeV. The lowest curve corresponds to zero temperature.
The dashed curve shows the boundary of phase transition liquid--gas, see the text.
}
%9
\label{f9}
\end{figure}
%%%%%%%%%%%%%%%%%%%%%%%%%%%%%%%%%%%%%%%%%%

It should, however, be recalled that these states of a Fermi-liquid are
unstable (it will be discussed below). It is hardly possible to apply directly the consideration
of zero sound given above to the situation of interacting quarks and antiquarks under studying,
because here the contribution of vacuum fluctuations of antiquarks, which form along with
quarks a chiral condensate, was completely ignored. On the other hand, zero sound
oscillations are known to be interpreted as a bound state of a particle and hole in the vicinity of the
Fermi sphere. Therefore, the excitations in a Fermi-liquid should be described (in our
case) by taking into account an interference between the bound states of a quark and antiquark, as
well as of a quark and a hole of the Fermi sphere (the quantum numbers of that hole allows
one to consider it as an antiparticle). We are doing that while calculating a respective
polarization operator.

Turning now to the chemical potential of quasiparticles presented in
Fig. \ref{f8} let us emphasize it is seen from the data for the NJL model that there is a region
of occupied states almost degenerate with respect to the chemical potential with the vacuum
chemical potential of a quasiparticle that quite naturally corresponds to the vanishing Fermi
momentum. Similarly, the chemical potential of occupied states in the KKB model differs from that
in vacuum by a small quantity proportional to the quark current mass
\begin{equation}
%35
\label{dmudro}
\frat{d \mu}{d \rho}=\frat{\mu}{\rho}\frat{v^2_{\mbox{\tiny{F}}}}{3}
\left(1+\frat{M_q}{P_{\mbox{\tiny{F}}}}\frat{d M_q}{d
P_{\mbox{\tiny{F}}}}\right)\sim m~.
\end{equation}
All the states with momentum $|{\bf p}|<2G$ are degenerate with respect to
the chemical potential in the chiral limit. $M_q=(4G^2-{\bf p}^2)^{1/2}$, $P_0=2G$, when
$P_{\mbox{\tiny{F}}}<|{\bf p}|<2G$, $M_q=0$,
$P_0=|{\bf p}|$ if $|{\bf p}|<P_{\mbox{\tiny{F}}}$, and $|{\bf p}|>2G$. Such
a behavior of the chemical potential is a consequence of a rapid decrease of the dynamical
quark mass with increasing Fermi momentum (see also (\ref{chem})). It follows from
Eq. (\ref{w2}) that the Fermi sphere is being filled as though from within. Those quarks with momenta
smaller than the Fermi one $|{\bf p}|<P_{\mbox{\tiny{F}}}$ do not take part in forming a condensate.
As a result, the quark dynamical mass can only decrease with the Fermi momentum increasing.
This dynamical mass is independent of the quark momentum in the NJL model because of the
approximation assumed. This dependence should be taken into account in more realistic case as an
analysis of the KKB model shows.

It comes about that the pressure of some occupied states degenerate in the
chemical potential almost coincides with that of vacuum (the pressure of a dilute Fermi-gas)
($T=0$)
$$P=-\frat{d E}{d V}=-{\cal E}+\mu~\rho~,$$
where ${\cal E}=E/V$ is the specific energy. Below we analyze respective data
in a more detail including the situation with nonzero temperature. The energy (and, hence, the
pressure) of ensemble is a discontinuous functional of the quark current mass (see
\cite{MZ}) in the KKB model. The integrands in (\ref{w2}) are estimated then as follows
$$p_0-P_0+\frat {1}{4G}~M^2 \sim -\frat{G~m^2}{p^2}~,$$
and we find a linearly diverging integral for the specific energy of ensemble
$$w\sim -\int \frat{dp~ p^2}{2\pi^2}~\frat{G~m^2}{p^2}~,$$
despite the fact that the delta-like form factor in the momentum space is the
strongest regularizer. It is paradoxical that any small value of the current mass $m$
leads to the negative infinite energy of ensemble, while the expression $w|_{m=0}$ is
well-defined in the chiral limit. Even more so, a similar divergence occurs in the case of a
delta-like form factor in the coordinate space. This fact is concealed by introducing the cutoff
momentum $\Lambda$ in the NJL model. Now it looks quite sensible to consider the relative
pressure of quark ensemble in comparison with a (formally infinite) vacuum value because of the
singular character (mentioned above) of ensemble pressure in the KKB model. The
pressure derivative in the ensemble density has the form: $d P/d \rho=\rho~d \mu/d \rho$. Therefore,
one can conclude by using an estimate given in (\ref{dmudro}) that the occupied states with
momenta $|{\bf p}|<2G$ are observed to degenerate with respect to the pressure
(${\cal E}=2G\rho$, $\mu=2G$) in the chiral limit in the KKB model. The deviations are
proportional to the quark current mass beyond the chiral limit.
%%%%%%%%%%%%%%%%%%%%%%%%%%%%%%%%%%%%%%%%%%%%%%%%%
\begin{figure}%[!tbh]
\includegraphics[width=0.3\textwidth]{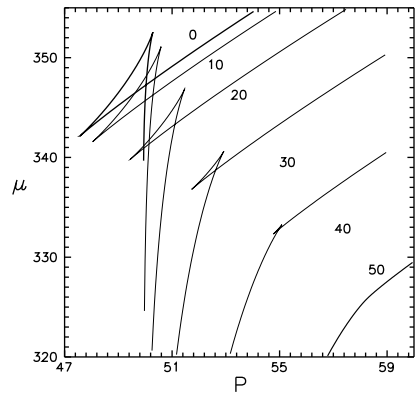}
\caption{The fragments of the isotherms shown in Fig. \ref{f9}, see the
text. The chemical potential $\mu$ (MeV) is plotted as a function of pressure
$P$ MeV/fm$^3$. The top curve corresponds to the zero isotherm and
following down with spacing $10$ MeV till the isotherm $50$ MeV (the lowest curve).
}
%10
\label{f10}
\end{figure}
%%%%%%%%%%%%%%%%%%%%%%%%%%%%%%%%%%%%%%%%%

Now, we are able to analyse some thermodynamic properties of a system and to
consider, first, the pressure of quark ensemble in detail
$$P=-\frat{d E}{d V}~.$$
By definition, the volume derivative should be calculated at the constant
mean entropy, $d\bar S/dV=0$. Implimanting this constraint, one can, for example, extract
the volume derivative of the chemical potential $d\mu/dV$. However, this approach cannot
be implied because mean charge conservation might be broken.
In fact, there is only one
possibility to satisfy both conditions by introducing two independent chemical potentials
for quarks and antiquarks separately. We use a symbol $\mu$ introduced earlier for the quark
chemical potential, whereas the antiquark chemical potential is taken with an opposite
charge and is denoted by $\bar\mu$. Then, we have
$$n=\frat{1}{e^{\beta~(P_0-\mu)}+1}~,~~\bar n=\frat{1}{e^{\beta~(P_0+\bar\mu)}+1}~$$
for the quark and antiquark densities, respectively. Some nonequilibrium
states of quark ensemble could also be described on this way (formally with a loss of
covariance, just similar to the electrodynamics as for the situation of electron—positron gas).
However, we are here interested only in a special configuration when $\bar\mu=\mu$. The partial
derivative of a specific energy $d w/d V$ can be presented in the following form
\begin{eqnarray}
%tv3
\label{tv3}
&&\frat{d w}{d V}=\int d \widetilde {\bf p}
\left(\frat{d n}{d\mu}\frat{d\mu}{dV}+\frat{d \bar n}{d\bar\mu}~
\frat{d\bar\mu}{dV}\right)
\biggl[p_0\cos\theta-2G \times\nonumber\\
&&\times\sin\left(\theta-\theta_m\biggr)
\int d \widetilde {\bf q}
~\sin\left(\theta'-\theta'_m\right)~(n'+\bar n'-1)~F\right]~.\nonumber
\end{eqnarray}
Dealing with the definition of an induced quark mass (\ref{9}) and presenting
the trigonometric factors via the quark dynamical mass we find out the ensemble pressure as
\begin{equation}
%36
\label{press}
P=-\frat{E}{V}-V~2N_c~\int d \widetilde {\bf p}~
\left(\frat{d n}{d\mu}~\frat{d\mu}{dV}+\frat{d \bar n}{d\bar\mu}~
\frat{d\bar\mu}{dV}\right)~P_0~.
\end{equation}
The condition of mean charge conservation
\begin{equation}
%37
\label{q4}
\frat{d \bar Q_0}{d V}=\frat{\bar Q_0}{V}+V~2N_c~\int d \widetilde {\bf p}~
\left(\frat{d n}{d\mu}~\frat{d\mu}{dV}-\frat{d \bar n}{d\bar\mu}~
\frat{d\bar\mu}{dV}\right)=0~,
\end{equation}
gives the first equation that interrelates the derivatives $d\mu/dV$ and
$d\bar\mu/dV$. Here, a regularized expression for the mean charge of quarks and antiquarks is
assumed modulo respective vacuum contribution.
%%%%%%%%%%%%%%%%%%%%%%%%%%%%%%%%%%%%%%%%%%%%%%%%%
\begin{figure}%[!tbh]
\includegraphics[width=0.3\textwidth]{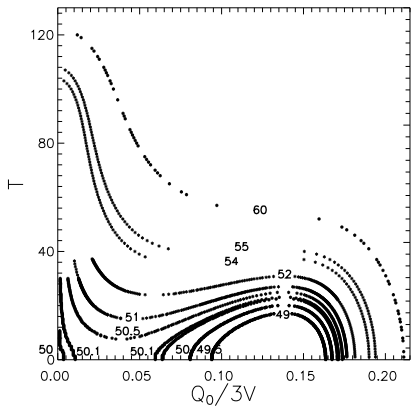}
\caption{Isobars of NJL model. Pressure in MeV/fm$^3$ is indicated next
to each curve. Vacuum pressure corresponds to approximately $50$ MeV/fm$^3$.}
%11
\label{f11}
\end{figure}
%%%%%%%%%%%%%%%%%%%%%%%%%%%%%%%%%%%%%%%%%

Implimenting the condition of constant mean entropy $d \bar S/dV=0$ in a
similar way one can obtain the second equation of chemical potential derivatives system as
follows
\begin{equation}
%38
\label{entr}
\hspace{-0.2cm}
\int \!\!\! d \widetilde {\bf p}\frat{d n}{d\mu}\ln \frat{n}{1-n}\frat{d\mu}{d V}
-\!\!\!\int\!\!\! d \widetilde {\bf p}\frat{d \bar n}{d\bar\mu}
\ln \frat{\bar n}{1-\bar n}\frat{d\bar\mu}{d V}=\frat{\bar S}{2N_cV^2}.
\end{equation}
Substituting the expressions $T~\ln\frat{n}{1-n}=\mu-P_0$ and
$T~\ln\frat{\bar n}{1-\bar n}=-\bar\mu-P_0$ into this equation and collecting
similar terms we come to the following equation
$$\int d \widetilde {\bf p}~
\left(\frat{d n}{d\mu}~\frat{d\mu}{dV}+\frat{d \bar n}{d\bar\mu}~
\frat{d\bar\mu}{dV}\right)~P_0=-\frat{\bar S~T}{2N_c~V^2}-\frat{\bar
Q_0~\mu}{2N_c~V^2}~.$$
if the condition $\bar\mu=\mu$ and Eq. (\ref{q4}) are satisfied. Finally, we
have for the pressure
\begin{equation}
%39
\label{p}
P=-\frat{E}{V}+\frat{\bar S~T }{V}+\frat{\bar Q_0~\mu}{V}~.
\end{equation}
Then the thermodynamic potential $\Omega$ should obey the following
thermodynamic identity
\begin{equation}
%40
\label{id}
\Omega=-P V=E-\mu~\overline{Q}_0 -T~\overline{S}~,
\end{equation}
as it should be/
At low temperatures the antiquark contribution is small and thermodynamic
description can be approximately developed by using the chemical potential $\mu$ only. If the
antiquark contribution becomes significant, a thermodynamic description is more
sophisticated and should obviously include the chemical potential $\bar\mu$ with additional condition
$\bar \mu = \mu$. Fig. \ref{f9} shows the ensemble pressure $P$ in MeV/fm$^3$ as a function of
the charge density ${\cal Q}_0/3V$ for various temperatures. The lowest curve is obtained at
zero temperature. Next curves following upwards correspond to temperatures $T = 10$ MeV, $T =
50$ MeV (an upper curve) with a step $T = 10$ MeV. Let us also remember the pressure of vacuum
for the NJL model was estimated in \cite{MZ} to be $40$ to $50$ MeV/fm$^3$ which is quite
consistent with that obtained in the bag model. It was also demonstrated that there is a region of
instability within a certain interval of the Fermi momenta generated by the anomalous behavior
of pressure $dP/dn<0$ (see also \cite{TM}). Fig. \ref{f10} displays fragments of isotherms
shown in Fig. \ref{f9} (but now in different coordinates) in the form of chemical
potential as a function of the ensemble pressure. A top curve is obtained at zero temperature. The
isotherms following below are shown in steps of $10$ MeV. A lowest curve is obtained at
temperature $50$ MeV. It is clearly seen from the figure that there are states on the isotherms which are
in thermodynamic equilibrium. The pressure and chemical potential are the same for these
states (see the characteristic Van der Waals triangle with intersecting curves). The
equilibrium points obtained
%%%%%%%%%%%%%%%%%%%%%%%%%%%%%%%%%%%%%%%%%%%%%%%%
\begin{figure}%[!tbh] \hspace{-0.0cm}
\includegraphics[width=0.3\textwidth]{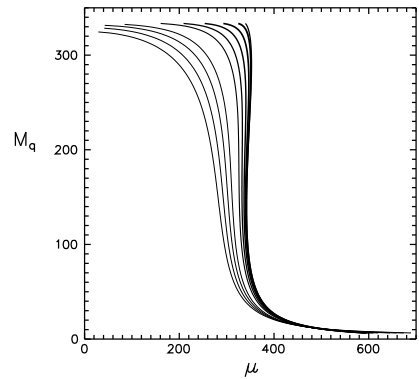}
\caption{The dynamical quark mass $|M_q|$ (MeV) as a function of chemical
potential $\mu$ (MeV) at the temperatures $T=0$ MeV, \dots , $T=100$ MeV with
spacing $T=10$ MeV. The most right curve corresponds to zero temperature.
}
%12
\label{ftv12}
\end{figure}
%%%%%%%%%%%%%%%%%%%%%%%%%%%%%%%%%%%%%%%%%%%%%%%%%
%%%%%%%%%%%%%%%%%%%%%%%%%%%%%%%%%%%%%%%%%%%%%%%%
are shown in Fig. \ref{f9} by a dashed curve. The points at which a dashed
curve intersects with isotherm give a boundary for a gas—liquid phase transition. The respective
line $P=\mbox{const}$ cuts off nonequilibrium and unstable fragments of isotherm and describes a
mixed phase. The critical temperature turns out to be equal to $T_c \approx 46$ MeV with the
critical charge density $\bar Q_0\approx 0.12$ charge/fm$^3$ for the above mentioned tuning
parameters. Fig. \ref{f11} shows the isobars. The pressure next to each curve is given in
MeV/fm$^3$. The vacuum pressure corresponds to approximately $\sim 50$ MeV/fm$^3$. It is
possible to extrapolate isobars into the region of small charge densities, however, it is not really
necessary. The figure clearly demonstrates the presence of dilute (a gas) and dense
(a liquid) phases in the vicinity of the vacuum isobar.

\begin{figure}%[!tbh] \hspace{-0.0cm}
\includegraphics[width=0.3\textwidth]{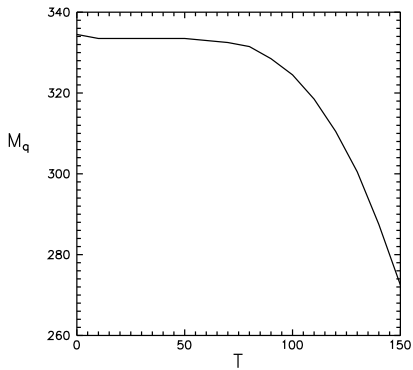}
\caption{ The dynamical quark mass $|M_q|$ (MeV) as a function of
temperature at the small value of charge density ${\cal Q}_4$.}
%13
\label{ftv12a}
\end{figure}

Fig. \ref{ftv12} shows the quark dynamical mass $M_q$ (in MeV) as a function
of chemical potential $\mu$ (in MeV) for temperatures $T = 0$ MeV, $T = 100$ MeV in steps
of $T = 10$ MeV. The rightmost curve corresponds to zero temperature. At low temperatures
below $50$ MeV the quark dynamical mass is a multivalued function of chemical potential.
Fig. \ref{ftv12a} shows the quark dynamical mass as a function of temperature at small charge density
${\cal Q}_0\sim 0$.  This picture is easily recognizable in context of the NJL model. It is the
latter that is implied in a scenario of chiral invariance restoration under extreme
temperatures higher than $100$ MeV and with a highly diluted quark ensemble. We have already noted
(see also \cite{MZ}) that the momentum $p_\theta$, which corresponds to the strongest
quark—antiquark attraction $d\sin \theta/d p=0$, can be determined. For example, for the NJL model this
parameter is equal to
\begin{equation}
%41
\label{tvpt}
p_\theta=(M_q~m)^{1/2}~.
\end{equation}
Its inverse value is given by the characteristic effective size of a
quasiparticle $r_\theta=p_\theta^{-1}$. From the behavior of the quark dynamical mass as a
function of temperature at small charge densities (see Fig. \ref{ftv12a}) one can
conclude that a quasiparticle size grows with an energy increasing.
%%%%%%%%%%%%%%%%%%%%%%%%%%%%%%%%%%%%%%%%%%%%%%%%
%%%%%%%%%%%%%%%%%%%%%%%%%%%%%%%%%%%%%%%%%%%%%%%%%

In \cite{MZ2} it was shown that if the quark chemical potential is defined as
energy necessary to add (remove) one quasiparticle, $\mu=dE/dN$, then the chemical potential
in vacuum coincides with the quark dynamical mass (see also (\ref{1var}), (\ref{chem})).
Therefore, it seems to be reasonable to consider a QCD phase diagram by starting from this value of the
chemical potential, though formally it can be taken smaller than the quark dynamical mass.
In particular, we exactly reproduce a standard
picture \cite{5}, \cite{ayz} by taking the chemical potential equal zero. The results obtained
allows to conjecture that the phase transition of (partial) restoration of the chiral
invariance could already be realized in nature as a mixed phase of physical vacuum and
baryonic matter. An indirect confirmation of this hypothesis can be seen in degenerate excited
states of some baryons (see, for instance, \cite{Gloz}). It is, however, clear that the data
presented (in particular, on the temperature and density of a critical point position)
should be understood as just an estimates. The critical temperature of a gas—liquid transition for
nuclear matter extracted from experiment is estimated to be about $20$ MeV. In addition,
here (at $T = 0$) a gas component possesses the nonzero density of order of $0.01$ of the normal
nuclear density, whereas an observed value should correspond to physical vacuum, i.e., to zero
baryon density. It should be noted that although such an uncertainty is inherent in the other
predictions of chiral symmetry restoration phase transition which are widely discussed in
many papers, they are somewhere around two to six normal nuclear matter densities.

\section{Polarization operator}
Returning to the discussion of zero sound and excitations of a chiral
condensate we would like also to remind that this knowledge is necessary for a more consistent
analysis of the transition gas—liquid layer. To this end, we will
need to know a polarization operator of the form
\begin{equation}
%28
\label{polop}
\Pi^{\Gamma}(p,q)=\int \frat{d k}{(2\pi)^4}~i~\pi^{\Gamma}(k+p,k-q)~,
\end{equation}
where
$$\pi^{\Gamma}(k+p, k-q)={\mbox {Tr}}\{S(k+p)\Gamma S(k-q)\Gamma\}~,$$
is a respective density of the polarization operator in the channels
$\Gamma=1, i\gamma_5,
\gamma_\mu, \gamma_5\gamma_\mu$ with the Green function of quark with the
dynamical mass $M_q$
\begin{equation}
%29
\label{gfun}
S(k)=\frat{1}{\hat k+\hat\mu-M_q({\vf k})}~,
\end{equation}
$\hat\mu=\mu \gamma^0$,
where $p$, $q$ are the incoming and outgoing external momenta of quark
quasiparticles. It will be enough for our purposes to consider the quasiparticles with momenta
${\vf p}={\vf q}={\vf Q}/2$ in the center of mass frame. We analyse
pseudoscalar and scalar
%%%%%%%%%%%%%%%%%%%%%%%%%%%%%%%%%%%%%%%%%%%%%%%%%
\begin{figure}%[!tbh]
\includegraphics[width=0.3\textwidth]{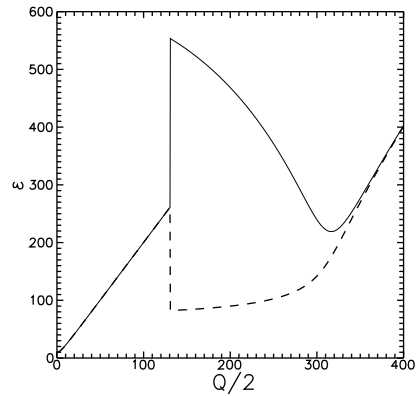}
\caption{
Energies (in MeV) of $\pi$- (dashed line) and $\sigma$- (solid line)
mesons as a function of momentum $Q/2$ (in MeV) ($T = 0$) for a gas of
low baryon density such that $P_{\mbox{\tiny{F}}}\sim 130$ MeV.
}
%14
\label{f12}
\end{figure}
%%%%%%%%%%%%%%%%%%%%%%%%%%%%%%%%%%%%%%%%%
channels, for which one can deduce
\begin{eqnarray}
%30
\label{pisi}
&&\hspace{-0.5cm}\Pi^{\pi,\sigma}=N_c\!\!\int d\widetilde
{\vf k}F({\vf k})
\left[\frat{a +b~\varepsilon}{\varepsilon^2-(E_{\vf p}+E_{\vf q})^2}+
\frat{c}{\varepsilon-E_{\vf p}+E_{\vf q}}\right],\nonumber\\[-.1cm]\\ [-
.1cm]
&&\hspace{-0.5cm}a\!\!=\!\!\biggl(E_{\vf p}+E_{\vf q}\biggr)\!\!
\biggl[2-n_{\vf p}-n_{\vf q}\biggr]\!\!
\left[\frat{{\vf Q}^2/4-{\vf k}^2\mp M_{\vf p} M_{\vf q}}
{E_{\vf p} E_{\vf q}}-1\right], \nonumber\\
&&\hspace{-0.5cm}b=\biggl[n_{\vf q}-n_{\vf p}\biggr]
\left[\frat{{\vf Q}^2/4-{\vf k}^2\mp
M_{\vf p} M_{\vf q}}
{E_{\vf p} E_{\vf q}}-1\right]~, \nonumber\\
&&\hspace{-0.5cm}c=[n_{\vf p}-n_{\vf q}]\left[\frat{{\vf Q}^2/4-{\vf
k}^2\mp
M_{\vf p} M_{\vf q}}
{E_{\vf p} E_{\vf q}}+1\right]~, \nonumber
\end{eqnarray}
where $\varepsilon=p_0-q_0$ is the transferred energy, $M_{\vf p}=M_q({\vf p})$,
$E_{\vf p}=[{\vf p}^2+M_q^2({\vf p})]^{1/2}$, the quantity $F({\vf k})$ is a
form factor and for the kinematics chosen ${\vf p}={\vf k}+{\vf Q}/2$ (then $n_{\vf p}$ is an
occupation number for a quasiparticle with momentum ${\vf p}$). In particular, at zero
temperature we have the Fermi step: $n_{\vf p}=n(E_{\vf p}-\mu)$. The similar notation is also
used for a quasiparticle with momentum ${\vf q}={\vf k}-{\vf Q}/2$.

\begin{figure}%[!tbh]
\includegraphics[width=0.3\textwidth]{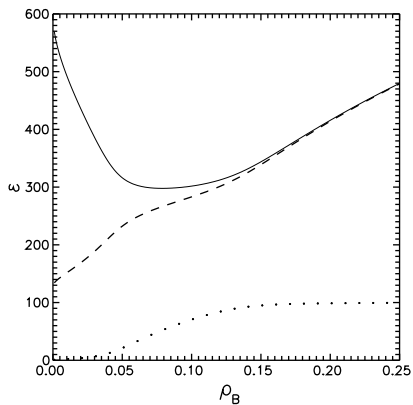}
\caption{Energies (in MeV) of $\pi$- (dashed line) and $\sigma$- (solid
line) mesons as a function of baryon density ($T=0$). The branch
corresponding to a bound quark-hole state is shown by dots $Q_q = -Q_h = 50$ MeV.
}
%15
\label{f13}
\end{figure}

The first term in Eq. (\ref{pisi}) corresponds to the quark and anti-quark
contributions, whereas the second one comes from the quark and hole configuration. It is easy to see
that at $F({\vf k})=\delta({\vf k})$ (in the KKB model) we have cubic dispersion
relations to determine the bound states: $1-2G~\Pi^{\pi,\sigma}(\varepsilon, {\vf Q})=0$. As an
example, Fig. \ref{f12} shows the energies (in units of MeV) calculated of $\pi$-(dashed line) and
$\sigma$-(solid line) mesons as functions of the momentum $Q/2$ (in units of MeV) at zero
temperature for the gas with small baryon density corresponding to the Fermi momentum
$P_{\mbox{\tiny{F}}}\sim 130$ MeV. The region of degeneracy, seen in Fig. \ref{f12} at low quark momenta, is a
consequence of the above discussed fact that the Fermi sphere is filled from the inside, and quarks
with momenta smaller than $P_{\mbox{\tiny{F}}}$ do not participate in forming the quark dynamical
mass. Such a behavior is not observed in the NJL model because of the approximation
adopted (the quark mass is independent of the momentum). Fig. \ref{f13} shows the energies (in units
of MeV) of $\pi$-(dashed line) and $\sigma$-(solid line) mesons as the functions of baryon
density ($T = 0$). The dots indicate a branch corresponding to the quark-hole bound state which
appears to be degenerate for $\pi$- and $\sigma$- mesons. Just these branches correspond to
the third additional root of the dispersion equation mentioned above, albeit there are
only two roots, (see discussion of the NJL model). To be specific, the quark momentum is
assumed to be $50$ MeV larger than the Fermi one but the hole momentum is $50$ MeV smaller than the
latter in this example. As it follows from Eq. (\ref{pisi}), the polarization operator in
the NJL model is defined by integrating over the running quark momentum $k$ and is represented
as a superposition of branches of the KKB model, which has already been mentioned in the
introduction. The most significant contributions (for the kinematics we chose) are those coming from
the terms denoted as $a$ and $c$ in Eq. (\ref{pisi}). Integrating over the angle (it is more
convenient to express the final formula by going to a nonsymmetric integration point,
the corrections become negligibly small) one can obtain (at $T=0$) the following results
\begin{eqnarray}
&&\hspace{-
0.5cm}\Pi^{\pi,\sigma}=A^{\pi,\sigma}+B^{\pi,\sigma}~,\nonumber\\
%[-.1cm]\\ [-.1cm]
&&\hspace{-
0.5cm}A^{\pi,\sigma}=\int\limits^{\Lambda_{P_{\mbox{\tiny{F}}}}}
\frat{d k~ k}{2\pi^2 Q}\left[\biggl(E_+-E_-\biggr)\left(1+\frat{E_++E_-}{2 E_k}\right)-
\right.\nonumber\\
&&\hspace{-0.6cm}\left.-\frat{Q^2-\varepsilon^2+2(M_q^2\mp M_q^2)}{2 E_k}
\ln \left(\!\!\frat{\varepsilon+E_k+E_+}{\varepsilon+E_k+E_-}
\frat{\varepsilon-E_k-E_+}{\varepsilon-E_k-E_-
}\!\!\right)\!\!\right]\!\!,\nonumber\\
&&\hspace{-0.5cm}B^{\pi,\sigma}=\int\limits_0^{P_{\mbox{\tiny{F}}}}
\frat{d k~ k}{2\pi^2 Q}\left[\biggl(E_+-E_-\biggr)
\left(1-\frat{E_++E_-}{2 E_k}\right)+\right.\nonumber\\
&&\hspace{-0.6cm}\left.+\frat{Q^2-\varepsilon^2+2(M_q^2\mp M_q^2)}{2 E_k}
\ln \left(\!\!\frat{\varepsilon-E_k+E_+}{\varepsilon-E_k+E_-}
\frat{\varepsilon+E_k-E_+}{\varepsilon+E_k-E_-}\!\!\right)
\!\!\right]\!\!,\nonumber
\end{eqnarray}
where $E_\pm=[(k\pm Q)^2+M_q^2]^{1/2}$ and $E_k=[k^2+M_q^2]^{1/2}$.

At small momentum $Q$ the component $B^{\sigma}$ is transformed into
Eq. (\ref{dzs}) with the parameter $s=E_{\mbox{\tiny{F}}}\varepsilon/(k_{\mbox{\tiny{F}}}Q)$. The
first component $A^{\pi,\sigma}$ results from the contribution coming from a quark—antiquark
pair, whereas the second one $B^{\pi,\sigma}$ arises due to the coupling of quark and hole
residing in the vicinity of the Fermi sphere. It should be noted that for a quark ensemble we
consider the medium properties which are mainly governed by the term $A^{\pi,\sigma}$
responsible for the quark—antiquark condensate, contrary to what we have in the condensed matter
physics where the dominant contribution, as it is known, is given by $B^{\pi,\sigma}$.
Therefore, the results obtained exclusively by using an analogy with the condensed matter physics
should be taken with a grain of salt. In particular, in the present paper we have analysed in
detail a situation with the zero sound description taken as an example illustrating just this point.
The zero sound would represent in itself the highly damped oscillations described by the
only scalar parameter $F_0$ while with no an antiquark presence. More accurate analysis shows that,
for example, there is a stable branch of quark and hole excitations in the Fermi sphere in
addition to a paired quark—antiquark state in the KKB model. We observe a regular mass convergence
for $\pi$- and $\sigma$-mesons when baryon density increases by performing numerical
integrationin the NJL model. This effect is clearly related to the restoration of chiral symmetry.
An influence of a bound quark—hole state in the Fermi sphere turns out to be insignificant. For
instance, for the densities of order of normal nuclear matter the dispersion law changes by a
few MeV when the quark and hole momentum differs more than $200$ MeV, but there are no damped
oscillations as it is in the KKB model.

One of the drawbacks of the models studied so far is the lack of quark
confinement that is understood here simply as an impossibility to observe a single particle state
with a regular (real) dispersion law. We see formally one quasiparticle can freely
propagate, indeed. But adding just another quasiparticle can dramatically change the picture due to
existence of a bound channel. For example, in the KKB model the bound states in scalar,
pseudoscalar, vector, and axial-vector channels appear at any quasiparticles momenta (details can
be found in \cite{wemes}). In particular, the bound state energy, obtained by using the
dispersion equation $1-2G~\Pi=0$, has the form
$$\varepsilon^2_{\pi,\sigma}\!\!=\!\!\biggl(E_{\vf p}+E_{\vf q}\biggr)^2\!\!-
2G\frat{E_{\vf p}+E_{\vf q}}
{E_{\vf p} E_{\vf q}}\!\!\biggl(E_{\vf p} E_{\vf q}\pm M_{\vf p} M_{\vf q}-{\bf
p}{\bf q}\biggr),$$
in $\pi$ and $\sigma$ channels (an upper sign corresponds to the pseudoscalar
channel). The first term in this expression is the energy of free particle motion. The
second one is strictly positive at any momenta $p$ and $q$ and plays a role of binding energy in
$\pi$ and $\sigma$ channels (only in the configuration of $q = p$ the binding energy vanishes
for a scalar channel). Similarly, one can show that a quark and antiquark are always coupled in
vector and axial—vector channels, i.e. the scattering matrix is always singular excepting a tensor
channel where it is trivial because of the initial interaction Hamiltonian that is taken as a
product of two color currents. Similar bound states exist in a diquark channel. As a consequence,
the states with any number of quark quasiparticles turn out to be the bound states in the
channels we have just mentioned. The same behavior is observed in the NJL model where the bound
states appear for the quarks with momenta somewhat lower than the cutoff momentum, i.e. the
scattering matrix is also singular within this momentum interval as in the KKB model. It seems the
bound states appear rather due to the fermion correlations than a physical influence of field
that is familiar in the quantum electrodynamics. Then, in order to understand what may take place
beyond the cutoff momentum one has apparently to study the appropriate nonlocal models.
%%%%%%%%%%%%%%%%%%%%%%%%%%%%%%%%%%%%%%%%%%%%%%%%%
%%%%%%%%%%%%%%%%%%%%%%%%%%%%%%%%%%%%%%%%%

\section{Transition layer between gas and liquid}
%II_1
The concept of a mixed phase of physical vacuum and baryonic matter would
receive the substantial confirmation if we are able to demonstrate an existence of the boundary
(transition) layer where a transformation of the quark ensemble from one aggregate state to another
takes place. As it was argued above the indicative characteristic to explore a homogeneous phase (at
finite temperature) is the mean charge (density) of ensemble. All the other characteristics, for
example, a chiral condensate, dynamical quark mass, etc. can be reconstructed if one knows the
ensemble mean charge. So, here we analyse a specific case of the surface (transition) layer
at zero temperature.

%II_2
We assume that the quark ensemble parameters in a gaseous phase are
approximately the same as those at zero charge $\rho_g=0$, i.e. as in vacuum (minor differences in
pressure, chemical potential and quark condensate are neglected). The dynamical quark mass
develops here the maximal value, and it is $M=335$ MeV for the parameter choice standard for the NJL
model. Then as the Van der Waals diagram shows a liquid phase, being in equilibrium with a gas
phase, gains the density $\rho_l=3\times 0.185$ charge/fm$^3$ (by some reason which becomes clear
below we correct it to take the value $\rho_l=3\times 0.157$ charge/fm$^3$). The detached factor $3$
here links again the magnitudes of quark and baryon matter densities. The quark mass is
approximately $\stackrel{*}{M}\approx 70$ MeV in this phase. Hereafter we focus on
describing two adjoining semi-infinite layers (i.e. assuming a plane symmetry of the corresponding
one-dimensional problem).

%II_3
The precursor experience teaches that an adequate description of
heterogeneous states can be reached with the mean field approximation \cite{Lar}. In our particular case
it means making use the corresponding effective quark-meson Lagrangian \cite{GL} (functional of
the Ginzburg-Landau type)
\begin{eqnarray}
\label{mesons}
&&{\cal L}=-\bar q~(\hat \partial +M)~q- \frat12~(\partial_\mu \sigma)^2-
U(\sigma)-\nonumber\\[-.2cm]
\\ [-.25cm]
&&-\frat14~F_{\mu\nu}F_{\mu\nu}-\frat{m_v^2}{2}~V_\mu V_\mu-
g_\sigma~\bar q q~\sigma+i g_v~\bar q~\gamma_\mu~q~ V_\mu~,\nonumber
\end{eqnarray}
where
$$F_{\mu\nu}=\partial_{\mu} V_\nu-\partial_{\nu} V_\mu~,~~~
U(\sigma)=\frat{m_\sigma^2}{2} ~\sigma^2+ \frat{b}{3}~\sigma^3
+\frat{c}{4}~\sigma^4~,$$
and $\sigma$ is the scalar field, $V_\mu$ is the field of vector mesons,
$m_\sigma$, $m_v$ are the masses of scalar and vector mesons and $g_\sigma$, $g_v$ are the coupling
constants of quark-meson interaction. The $U(\sigma)$ potential includes the nonlinear
$\sigma$ field interaction terms up to the fourth order, for example. For the sake of
simplicity we do not includ the contributions coming from the pseudoscalar and axial-vector
mesons.

%II_4
The meson component of such a Lagrangian should be self-consistently treated
by considering the corresponding quark loops. Here we do not see any reason to go beyond the
well elaborated and reliable one loop approximation (\ref{mesons}) \cite{GL}, although recently
the considerable progress has been reached  in
scrutinizing the non-homogeneous quark condensates by applying the powerful methods of exact
integration \cite{KN}. Here we believe it is more practical to adjust phenomenologically
the effective Lagrangian parameters basing on the transparent physical picture. It is easy
to see that handling (\ref{mesons}) one loop approximation we come, in actual fact, to the Walecka
model \cite{wal} but adopted for the quarks. In what follows we are working with the
designations of that model and do hope it does not lead to the misunderstandings.

%II_5
In the context of our paper we propose to interpret Eq. (\ref{mesons}) in the
following way. Each phase might be considered, in a sense, with regard to another phase as
an excited state which requires the additional (apart from a charge density) set of parameters
(for example, the meson fields) for its complete description, and those are characterizing the
measure of deviation from the equilibrium state. Then the crucial question becomes whether it is
possible to adjust the parameters of effective Lagrangian (\ref{mesons}) to obtain the solutions
in which the quark field interpolates between the quasiparticles in the gas (vacuum) phase and
the quasiparticles of the filled-up states. For all that the density of the filled-up state
ensemble should asymptotically approach the equilibrium value of $\rho_l$ and should turn to
the zero value in the gas phase (vacuum).

%II_6
The scale inherent in this problem may be assigned with one of the mass
referred in the Lagrangian (\ref{mesons}). In particular, we bear in mind the dynamical quark
mass in the vacuum $M$. Besides, there are another four independent parameters in the problem
and in order to compare them with the results of studying a nuclear matter we employ the form
characteristic for the (nuclear) Walecka model
$$C_s=g_\sigma~\frat{M}{m_\sigma}~,~~C_v=g_v~\frat{M}{m_v}~,
~~\bar b=\frat{b}{g_\sigma^3~M}~,~~\bar c=\frat{c}{g_\sigma^4}~.$$
Parameterizing the potential $U(\sigma)$ as
$b_\sigma=1.5~m_\sigma^2~(g_\sigma/M)$,
$c_\sigma=0.5~m_\sigma^2~(g_\sigma/M)^2$ we come to the sigma model whereas
the choice $b=0$, $c=0$ results in the Walecka model. As to standard nuclear matter application
the parameters $b$ and $c$ demonstrate vital model dependent character and are quite
different from the parameter values of sigma model. Truly, in that case their values are also
regulated by additional requirement of an accurate description of the saturation property.
On the other hand, for the quark Lagrangian (\ref{mesons}) we could intuitively anticipate some
resemblance with the sigma model and, hence, could introduce two dimensionless parameters
$\eta$ and $\zeta$ in the form of $b=\eta~b_\sigma$, $c=\zeta^2~c_\sigma$ which characterize some
fluctuations of the effective potential. Then the scalar field potential is presented as follows
$$U(\sigma)=\frat{m_\sigma^2}{8}~\frat{g_\sigma^2}{M^2}~
\left(4~\frat{M^2}{g_\sigma^2}+4~\frat{M}{g_\sigma}~\eta~\sigma+\zeta^2\sigma^2\right)~\sigma^2~.$$

%II_7
The meson and quark fields are determined by solving the following system of
the stationary equations
\begin{eqnarray}
\label{sys}
&&\Delta~ \sigma -
m_\sigma^2~\sigma=b~\sigma^2+c~\sigma^3+g_\sigma~\rho_s~,\nonumber\\
&&\Delta ~V - m_v^2~ V=-g_v~\rho~,\\
&&( {\bf \hat\nabla}+\stackrel{*}{M})~q=(E-g_v~V)~q~,\nonumber
\end{eqnarray}
where $\stackrel{*}{M}=M+g_\sigma\sigma$ is the running value of dynamical
quark mass, $E$ stands for the quark energy and $V=-iV_4$. The density matrix describing the quark
ensemble at $T=0$ has the form
$$ \xi (x)=\int\limits^{P_F} d \widetilde{\bf p}~q_{\bf p}(x)~
\bar q_{\bf p}(x)~,$$
in which ${\bf p}$ is the quasiparticle momentum and the Fermi momentum $P_F$
is defined by the corresponding chemical potential. The densities $\rho_s$ and $\rho$ at the
right hand sides of Eq. (\ref{sys}) are by definition
$$\rho_s(x)=Tr\left\{\xi(x),1
\right\}~,~~\rho(x)=Tr\left\{\xi(x),\gamma_4\right\}~.$$

%II_8
Here we confine ourselves to the Thomas--Fermi approximation while describing
the quark ensemble. Then the densities which we are interested in are given with some local Fermi
momentum $P_F(x)$ as
\begin{eqnarray}
\label{rhorhos}
&&\hspace{-0.5cm}\rho= \gamma\int\limits^{P_F}d \widetilde {\bf p}=
\frat{\gamma}{6\pi^2}~P_F^3~,\\
&&\hspace{-0.5cm}\rho_s=\gamma\int\limits^{P_F}d \widetilde {\bf p}~
\frat{\stackrel{*}{M}}{E}=\nonumber\\
&&\hspace{-0.5cm}=\frat{\gamma}{4\pi^2}\stackrel{*}{M}P_F^2
\left\{\biggl(1+\lambda^2\biggr)^{1/2}\!\!\!-\frat{\lambda^2}{2}
\ln\left[\frat{\left(1+\lambda^2\right)^{1/2}+1}
{\left(1+\lambda^2\right)^{1/2}-1}\right]\right\},\nonumber
\end{eqnarray}
where $\gamma$ is a quark gamma-factor which for one flavour is $\gamma=2N_c$,
$E\!=\!({\bf p}^2+\stackrel{*}{M}^2\!)^{1/2}$ and
$\lambda=\stackrel{*}{M}/P_F$. Under assumption adapted the ensemble chemical potential is constant and,
therefore, a local value of the Fermi momentum is defined by the running value of dynamical
quark mass and vector field as
\begin{equation}
\label{mu}
\mu=M=g_v~V+\left(P_F^2+\stackrel{*}{M}^2\right)^{1/2}~.
\end{equation}

%II_9
Now we should tune the Lagrangian parameters in Eq. (\ref{mesons}). For
asymptotically large distances (in a homogeneous phase) we may neglect the gradients of scalar and
vector fields and the equation for scalar field of the system (\ref{sys}) leads to the first
equation that relates the parameters $C_s$, $C_v$, $\bar b$, $\bar c$ as
\begin{equation}
\label{id1}
\hspace{-0.1cm}
\frat{\stackrel{}{M}^2 \!\!\!\!(\stackrel{*}{M}-
\stackrel{}{M})}{C_s^2}+\bar b\stackrel{}{M}\!\!\!(\stackrel{*}{M}-
\stackrel{}{M})^2
+\bar c(\stackrel{*}{M}-\stackrel{}{M})^3=-\rho_s.
\end{equation}
The vector field asymptotically is given by the ensemble density
$V=C_v^2~\rho/(g_v M^2)$. The second equation derived from the relation (\ref{mu}) for the chemical
potential looks like
\begin{equation}
\label{id2}
M=\frat{C_v^2~\rho}{M^2}+\left(P_F^2+\stackrel{*}{M}^2\right)^{1/2}~.
\end{equation}

%II_10
If we know the liquid density we obtain the Fermi momentum ($P_F=346$ MeV)
from (\ref{rhorhos}) and applying the identities (\ref{id1}), (\ref{id2}) we have for the
particular case $b=0$, $c=0$ that $C_s^2=25.3$, $C_v^2=-0.471$, i.e. the vector component $C_v^2$ is small
(compared to $C_s^2$) and acquires a negative value that is unacceptable. Apparently, it
looks necessary to abandon the contribution coming from the vector field or to reduce the
dynamical quark mass $\stackrel{*}{M}$ up to the value which retains the identity (\ref{id2})
valid with positive $C_v^2$ or even zero value. In the gaseous phase the dynamical quark mass can
also be corrected to the value larger than the vacuum value. It is clear that in the situation
of the liquid with the density $\rho_l=3\times 0.185$ ch/fm$^3$ the dynamical quark mass should
coincide (or exceed) $M=346$ MeV in the gaseous phase. However, here we correct the liquid density
(as it was argued above) to decrease its value up to $\rho_l=3\times 0.157$ch/fm$^3$ which is
quite acceptable in the nucleation capacity. In fact, this possibility can be simply justified by
another choice of the NJL model parameters. Thus, we obtain at $\stackrel{*}{M}=70$ MeV and
$b=0$, $c=0$ that $C_s^2=28.4$, $C_v^2=0.015$, i.e. we have a small but positive value for the
vector field coefficient. At the same time, being targeted here to estimate the surface
tension effects only we do not strive for the precise fit of parameters. In the Walecka model
these coefficients are $C_s^2=266.9$, $C_v^2=145.7$, ($b=0$, $c=0$). Moreover, there is another
parameter set with $C_s^2=64.$, $C_v^2\approx 0$ \cite{Bog} but it is rooted in an essential
nonlinearity of the sigma-field due to the nontrivial values of the coefficients $b$ and $c$. The
option (formally unstable) with negative $c$ ($b$) has been also discussed.

%%%%%%%%%%%%%%%%%%%%%%%%%%%%%%%%%%%%%%%%%%%%%%%%%
\begin{figure}%[!tbh]
\begin{center}
\leavevmode
\hspace{-1.5cm}
\includegraphics[width=0.3\textwidth]{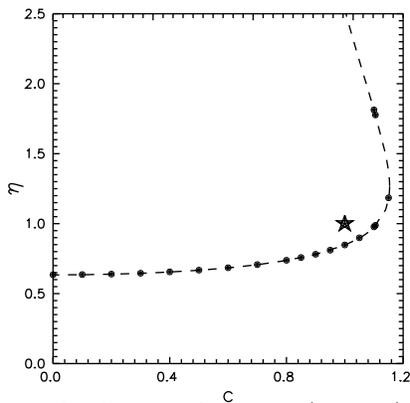}
\vspace{-5mm}
\caption{The domain of the $\eta$, $c$ ($\zeta=c~\eta$)-plot in which an
increase of specific energy occurs, see the text. The dots represent a
stable kink. The star shows the position of canonical (chiral) kink, see the text.
}
%16
\label{f5i}
\end{center}
\end{figure}
%%%%%%%%%%%%%%%%%%%%%%%%%%%%%%%%%%%%%%%%%%%%%%%%%

%II_11
The coupling constant of scalar field is fixed by the standard (for the NJL
model) relation between the quark mass and the $\pi$-meson decay constant $g_\sigma=M/f_\pi$
(we put $f_\pi=100$ MeV) although there is no any objection to treat this coupling constant as an
independent parameter. As a result of all agreements done we have for the $\sigma$-meson
mass $m_\sigma=g_\sigma~M/C_s$. In principle, we could even fix the $\sigma$-meson
mass and coupling constant $g_\sigma$ but all relations above mentioned lead eventually to
quite suitable values of the $\sigma$-meson mass as will be demonstrated below. The vector field
plays, as we see, a secondary role because of the small magnitude of constant $C_v$. Then taking
the vector meson mass as $m_v\approx 740$ MeV (slightly smaller value than the mass of
$\omega$-meson because of simple technical reason only) we calculate the coupling constant of vector
field from the relation similar to the scalar field $m_v=g_v~M/C_v$. Amazingly, its value
comes about steadily small being compared to the value characteristic for the NJL model
$g_v=\sqrt{6}g_\sigma$. However, at the values of constant $C_v$ which we are interested in it is
very difficult to maintain the reasonable balance and to be specific in this paper we prefer to
choose the massive vector field. Actually, it is unessential because we need this parameter (as
we remember) to estimate the vector field strength only.

%%%%%%%%%%%%%%%%%%%%%%%%%%%%%%%%%%%%%%%%%%%%%%%%%
\begin{figure}%[!tbh]
\begin{center}
\leavevmode
\includegraphics[width=0.3\textwidth]{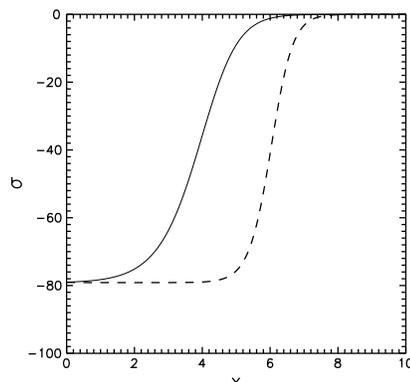}
\vspace{-5mm}
\caption{The stable kink solutions with $c=1.1$, the solid line
corresponds to $\eta\approx 0.977$ ($m_\sigma\approx 468$ MeV) and the
dashed line corresponds to $\eta\approx 1.813$ ($m_\sigma\approx 690$ MeV),
$x$ is given in the units of $fm$ and $\sigma$ is given in MeV.
}
%17
\label{f5a}
\end{center}
\end{figure}
%%%%%%%%%%%%%%%%%%%%%%%%%%%%%%%%%%%%%%%%%%%%%%%%%
%II_12
The key point of our interest here is the surface tension coefficient
\cite{Bog} which can be defined as
\begin{equation}
\label{sur}
u_s=4\pi~r_o^2~\int\limits_{-\infty}^{\infty}dx~
\left[ {\cal E}(x)-\frat{{\cal E}_l}{\rho_l}~\rho(x)\right]~.
\end{equation}
The parameter $r_o$ will be discussed in the next section at considering the
features of quark liquid droplet, and for the present we would like to notice only that for the
parameters considered its magnitude for $N_f=1$ is around $r_o=0.79$ fm. Recalling the
factor $3^{1/3}$ which connects the baryon and quark numbers, we find the magnitude
($\widetilde r_o=3^{1/3} 0.79\approx1.14$ fm) in full agreement with the
magnitude standard for the nuclear matter calculations (in the Walecka model)
$\widetilde r_o=1.1$---$1.3$ fm.

%II_13
In order to proceed we calculate ${\cal E}(x)$ in the Thomas--Fermi
approximation as
\begin{eqnarray}
{\cal E}(x)&=&\gamma\!\!\int\limits^{P_F(x)}d \!\!\widetilde {\bf p}~
[{\bf p}^2+\stackrel{*}{M}(x)]^{1/2}+\nonumber\\
&+&\frat12~g_v~\rho(x)~V(x)-
\frat12~g_\sigma~\rho_s(x)~\sigma(x)~.\nonumber
\end{eqnarray}
And to give some idea for the 'setup' prepared we present here the
characteristic parameter values for some fixed $b$ and $c$ with $\rho_l=3\times 0.157$ ch/fm$^3$. In
the liquid phase they are $\stackrel{*}{M}=70$ MeV ($P_F=327$ MeV) and $e_l=310.5$ MeV (index
$l$ stands for a liquid phase and $e(x)={\cal E}(x)/\rho(x)$ defines the density of specific
energy). Both relations (\ref{id1}) and (\ref{id2}) are obeyed by this state. There exist
the solution with larger value of quark mass $\stackrel{*}{M}=306$ MeV, ($P_F=135$ MeV) (we
have faced the similar situation in the first section dealing with the gas of quark quasiparticles)
and $e=338$ MeV $\sim e_g$ ($e_g$ is the specific energy in the gas phase) that
satisfies both equations as well. The specific energy of this solution occurs to be larger
than specific energy of the previous solution. It is also worthwhile to mention the existence of
intermediate state corresponding to the saturation point with the mass $\stackrel{*}{M}=95$ MeV,
($P_F=291$ MeV) and $e=306$ MeV. Obviously, it is the most favorable state with the smallest
value of specific energy (and with the zero pressure of quark ensemble), and the system can
fall into this state only in the presence of significant vector field. This state (already
discussed in the first section) corresponds to the minimal value of chemical potential ($T=0$) and
can be reached at the densities typical for the normal nuclear matter. However, Eq. (\ref{id2}) is
not valid for this state.
%%%%%%%%%%%%%%%%%%%%%%%%%%%%%%%%%%%%%%%%%%%%%
%\begin{figure*}[!tbh]
\begin{figure}[!tbh]
\begin{center}
\leavevmode
%\vspace{-4cm}
\includegraphics[width=0.3\textwidth]{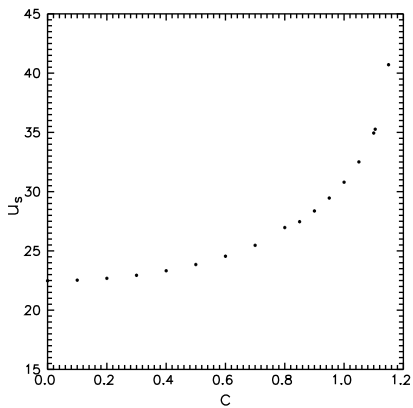}
\caption{The surface tension coefficient $u_s$ in MeV as a function
of parameter $c$ ($\zeta=c~\eta$) for the curve of stable kinks (with
$\eta\leq 1.2$).}
%18
\label{f6i}
\end{center}
\end{figure}
%\end{figure*}
%%%%%%%%%%%%%%%%%%%%%%%%%%%%%%%%%%%%%%%%%%%%%

%II_14
Two another parameters $\eta$, $\zeta$ are fixed by looking through all the
configurations in which the solution of equation system (\ref{sys}) with stable kink of the
scalar field does exist and describes the transition of quarks from the gaseous phase to the liquid
one. First, it is reasonable to scan the $\eta$, $c$ ($\zeta=c~\eta$)-plane, in order to
identify the domain in which the increase of specific energy ${\cal E}-{\cal E}_l~\rho/\rho_l\le 0$
is revealed at running through all possible states which provide the necessary transition
(without taking into account the field gradients). In practice one need to follow a simple
heuristic rule. The state with $P_F \sim 1$ MeV (i.e. $e$ and the corresponding $\rho$) and the state
of characteristic liquid energy ${\cal E}_l$ (together with $\rho_l$) should be compared while
scanning the Lagrangian parameters $\eta$ and $c$. Just the domain in which they are
commensurable could provide us with the solutions which we are interested in and Fig. \ref{f5}
shows its boundary. The curve could be continued beyond the value $\eta=2.5$ but the values of
corresponding parameter $\eta$ are unrealistic and not shown in the plot.

%II_15
We calculate the solution of equation system (\ref{sys}) numerically by the
Runge--Kutta method with the initial conditions $\sigma(L)\approx 0$, $\sigma'(L)\approx 0$
imposed at the large distance $L\gg t$, where $t$ is a characteristic thickness of transition
layer (about 2 fm). Such a simple algorithm occurs quite suitable if the
vector field contribution is considered as a small correction (what just takes place
in the situation under consideration) and is presented as
$$V(x)=\frat{1}{2m_v}~\int\limits^L_{-L}~dz~e^{-m_v|x-z|}~g_v~\rho(z)~,$$
where the charge (density) $\rho$ is directly defined by the scalar field. We
considered the solutions including the contribution of the vector field as well and the
corresponding results confirm the estimates obtained.

%II_16
Rather simple analysis shows the interesting solutions are located along the
boundary of discussed domain. Some of those are depicted in Fig. \ref{f5i} as the dots.
Fig. \ref{f5a} shows the stable kinks of $\sigma$-field with the parameter $c=1.1$ for two
existing solutions with
$\eta\approx 0.977$ ($m_\sigma\approx 468$ MeV) (solid line) and
$\eta\approx 1.813$ ($m_\sigma\approx 690$ MeV) (dashed line). For the sake
of clarity we consider the gas (vacuum) phase is on the right. Then the asymptotic value of
$\sigma$-field on the left hand side ($\sigma\approx 80$ MeV) corresponds to
$\stackrel{*}{M}=70$ MeV. The thickness of transition layer for the solution with $\eta\approx 0.977$ is
$t\approx 2$ fm whereas for the second solution $t\approx 1$ fm.

%II_17
Characterizing the whole spectrum of the solutions obtained we should mention
that there exist another more rigid (chiral) kinks which correspond to the transition into the
state with the dynamical quark mass changing its sign, i.e. $M\to-M$. In particular, the
kink with the canonical parameter values $\eta=1$, $c=1$ is clearly seen (marked by the star in Fig.
\ref{f5i}) and its surface tension coefficient is about $2m_\pi$ ($m_\pi$ is the $\pi$-meson
mass). The most populated class of solutions consists of those having the meta-stable
character. The system comes back to the starting point (after an evolution) pretty rapidly, and usually
the $\sigma$-field does not evolve in such an extent to reach the asymptotic value (which
corresponds to the dynamical quark mass in the liquid phase $\stackrel{*}{M}=70$ MeV). Switching
on the vector field changes the solutions insignificantly (for our situation with small $C_v$ it
does not exceed $2$ MeV in the maximum).

%%%%%%%%%%%%%%%%%%%%%%%%%%%%%%%%%%%%%%%%%%%%%%%%%
\begin{figure}%[!tbh]
\begin{center}
\leavevmode
\hspace{-1.5cm}
\includegraphics[width=0.3\textwidth]{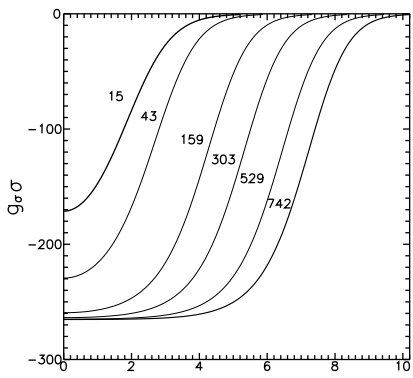}
\vspace{-5mm}
\caption{$\sigma$-field (MeV) as a function of the distance $r$ (fm) for
several solutions of the equation system (\ref{sys}) which are characterized
by the net quark number $N_q$ written down to the left of each curve.
}
%19
\label{f7i}
\end{center}
\end{figure}
%%%%%%%%%%%%%%%%%%%%%%%%%%%%%%%%%%%%%%%%%%%%%%%%%
%II_18
The surface tension coefficient $u_s$ in MeV for the curve of stable kinks
with parameter $\eta\leq 1.2$ as the function of parameter $c$ ($\zeta=c~\eta$) is depicted
in Fig. \ref{f6i}. The $\sigma$-meson mass at $c\approx 0$ is $m_\sigma\approx 420$ MeV and
changes smoothly up to the value $m_\sigma\approx 500$ MeV at $c\approx 1.16$ (the maximal value of
the coefficient $c$ beyond which the stable kink solutions are not observed). In particular,
$m_\sigma\approx 450$ MeV at $c=1$. Two kink solutions with $c=1.1$ for $\eta \approx 0.977$ and
for $\eta\approx 1.813$ (shown in Fig. \ref{f5a}, and the second one is not shown
in Fig. \ref{f6i}) have the tension coefficient values $u_s\approx35$ MeV and $u_s\approx 65$
MeV, correspondingly. The maximal value of tension coefficient for the normal nuclear matter does
not exceed $u_s=50$ MeV. The nuclear Walecka model claims the value $u_s\approx 19$ MeV
\cite{Bog} as acceptable and calculable. The reason to have this higher value of surface tension
coefficient for quarks is rooted in the different values of the mass deficit. Indeed, for nuclear
matter it does not exceed $\stackrel{*}{M}\approx0.5 M$ albeit more realistic values are considered
around $\stackrel{*}{M}\approx 0.7 M$ and for the quark ensemble the mass deficit
amounts to $\stackrel{*}{M}\approx 0.3 M$. We are also able to estimate the compression
coefficient of quark matter $K$ which occurs significantly larger than the nuclear one. Actually,
we see quite smooth analogy between the results obtained and the results of bag soliton model
\cite{sbm}. The thermodynamic treatment developed in the present paper allows us to formulate
the adequate boundary conditions for the bag in physical vacuum and to diminish
considerably the uncertainties in searching the true soliton Lagrangian. We believe, it was also shown here,
that to single out one soliton solution among others (including even those obtained by the exact
integration method \cite{KN}), which describes the transitional layer between two media, is not
easy problem if the boundary conditions above formulated are not properly imposed.

%%%%%%%%%%%%%%%%%%%%%%%%%%%%%%%%%%%%%%%%%%%%%%%%%%
% Section III
%%%%%%%%%%%%%%%%%%%%%%%%%%%%%%%%%%%%%%%%%%%%%%%%%%
\section{Droplet of quark liquid}
%III_1
The results of previous sections have led us to put the challenging question
about the creation and properties of finite quark systems or the droplets of quark liquid which
are in equilibrium with the vacuum state. Thus, as a droplet we imply the spherically symmetric
solution of the equation system (\ref{sys}) for $\sigma(r)$ and $V(r)$ with the obvious
boundary conditions $\sigma'(0)=0$ and $V'(0)=0$ in the origin (the primed variables denote the
first derivatives in $r$) and rapidly decreasing at the large distances $\sigma\to 0$,
$V\to 0$, when $r\to\infty$.

%III_2
A quantitative analysis of similar nuclear physics models which includes the
detailed tuning of parameters is usually based on the comprehensive fitting of available
experimental data. This way is obviously irrelevant in studying the quark liquid droplets. This
global difficulty dictates a specific tactics of analyzing. We propose to start, first of all,
with selecting the parameters which could be worthwhile to play a role of physical observables.
Naturally, the total baryon number which phenomenologically (via factor $3$) related to the number
of valence quark in an ensemble is a reasonable candidate for this role. Besides, the density of
quark ensemble $\rho(r)$, the mean size of droplet $R_0$ and the thickness of surface layer
$t$ look like suitable for such an analysis.

%III_3
It is argued above that the vector field contribution is negligible because
of the small value of coefficient $C_v$ compared to the $C_s$ magnitude, and we follow this
conclusion (or assumption) albeit understand it is scarcely justified in the context of
finite quark system. Thus, we put down $g_v=0$, $V=0$ in what follows and it simplifies all the
calculations enormously.

%%%%%%%%%%%%%%%%%%%%%%%%%%%%%%%%%%%%%%%%%%%%%%%%%
\begin{figure}%[!tbh]
\begin{center}
\leavevmode
\includegraphics[width=0.3\textwidth]{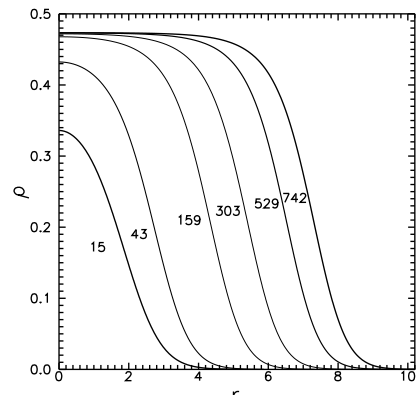}
\vspace{-5mm}
\caption{
Distribution of the quark density $\rho$ (ch/fm$^3$) for the
corresponding solutions presented in Fig. \ref{f7a}.
}
%20
\label{f7a}
\end{center}
\end{figure}
%%%%%%%%%%%%%%%%%%%%%%%%%%%%%%%%%%%%%%%%%%%%%%%%%
%III_4
Fig. \ref{f7i} shows the number of solutions ($\sigma$-field in MeV) to the
system (\ref{sys}) at $N_f=1$ and Fig. \ref{f7a} presents the corresponding distributions of
ensemble density $\rho$ (ch/fm$^3$). The parameters $C_s$, $C_v$, $b$ and $c$ are derived by
the same algorithm as in the previous section, i.e. the chemical potential of quark ensemble
$M=335$ MeV (and $\sigma \to 0$) is fixed at the spatial infinity. The filled-up states (of a
liquid) are characterized by the parameters $\stackrel{*}M=70$ MeV,
$\rho_0=\rho_l=3\times 0.157$ ch/fm$^3$.
The $\sigma$-meson mass and the coupling constant $g_\sigma$ are derived at
fixed coefficients $\eta$ and $\zeta$, and they just define the behaviour of solutions
$\sigma(r)$, $\rho(r)$, etc. The magnitudes of functions $\sigma(r)$ and $\rho(r)$ at origin are not
strongly correlated with the values characteristic for the filled-up states and are practically
determined by solving the boundary value problem for system (\ref{sys}). In particular, the solutions
presented in Fig. \ref{f7i} have been received with the running coefficient $\eta$ at
$\zeta=\eta$. The most relevant parameter (instead of $\eta$) from the physical view point is the
total number of quarks in the droplet $N_q$ (as discussed above) and it is depicted to the left of
each curve. (The variation of $\stackrel{*}M$, $\rho_0$ and $f_\pi$ could be considered as
well instead of two mentioned parameters $\eta$ and $\zeta$.)

%III_5
Analyzing the full spectrum of solutions obtained by scanning one can reveal
a recurrent picture (at a certain scale) of kink-droplets which are easily parameterized by the
total number of quarks $N_q$ in a droplet and by the density $\rho_0$. These characteristics
are obviously fixed at completing the calculations. The sign which allows us to single out these
solutions is related to the value of droplet specific energy (see below).
%%%%%%%%%%%%%%%%%%%%%%%%%%%%%%%%%%%%%%%%%%%%%
%\begin{figure*}[!tbh]
\begin{figure}[!tbh]
\begin{center}
\leavevmode
%\vspace{-4cm}
\includegraphics[width=0.25\textwidth]{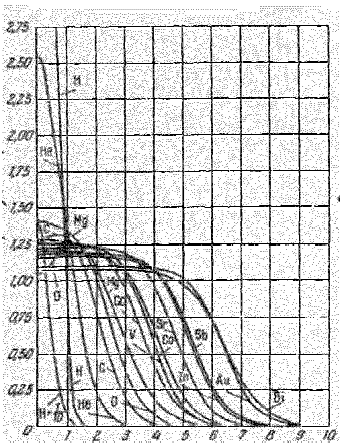}
\caption{Measured nuclear charge distributions (Hofstadter).}
%ad1
\label{fad1}
\end{center}
\end{figure}
%\end{figure*}
%%%%%%%%%%%%%%%%%%%%%%%%%%%%%%%%%%%%%%%%%%%%%

%III_6
Table \ref{tab:table1} exhibits the results of fitting the density $\rho(r)$
with the Fermi distribution
\begin{equation}
\label{ferdis}
\rho_F(r)=\frat{\widetilde \rho_0}{1+e^{(R_0-r)/b}}~,
\end{equation}
where $\widetilde \rho_0$ is the density in origin, $R_0$ is the mean size of
droplet and the parameter $b$ defines the thickness of surface layer $t=4\ln(3) b$. Besides,
the coefficient $r_0$ which is absorbed in the surface tension coefficient (\ref{sur}), the
$\sigma$-meson mass, $R_0=r_0 N_q^{1/3}$ and the coefficient $\eta$ at which all other values have
been obtained are also presented in the Table \ref{tab:table1}.
\begin{table}
\caption{\label{tab:table1}Results of fitting by the Fermi distribution
with $N_f=1$, $\widetilde\rho_0$ (ch/fm$^3$), $R_0$, $t$, $r_0$ (fm),
$b$ (fm$^{-1})$, $m_\sigma$ (MeV).}
\begin{ruledtabular}
\begin{tabular}{cccccccc}
$N_q$ &$\widetilde
       \rho_0$
                  &$R_0$
                          &$b$
                                     &$t$      &$r_0$     &$m_\sigma$
&
$\eta$ \\\hline
$15$  &$0.34$     &$1.84$ &$0.51$    &$2.24$   &$0.74$    &$351$
&
$0.65$    \\
$43$  &$0.43$     &$2.19$ &$0.52$    &$2.28$   &$0.75$    &$384$
&
$0.73$    \\
$159$ &$0.46$     &$4.19$ &$0.52$    &$2.29$   &$0.77$    &$409$
&
$0.78$    \\
$303$ &$0.47$     &$5.23$ &$0.52$    &$2.29$   &$0.78$    &$417$
&
$0.795$    \\
$529$ &$0.47$     &$6.37$ &$0.52$    &$2.27$   &$0.79$    &$423$
&
$0.805$    \\
$742$ &$0.47$     &$7.15$ &$0.52$    &$2.27$   &$0.79$    &$426$
&
$0.81$ \\
\end{tabular}
\end{ruledtabular}
\end{table}

%III_7
The curves plotted in the Fig. \ref{f7i} and results of Table
\ref{tab:table1} justify to conclude that the density distributions at $N_q\ge 50$ are in full agreement
with the corresponding data typical for the nuclear matter. The thicknesses of
transition layers in both cases are also similar and the coefficient $r_0$ with the factor $3^{1/3}$
included is in full correspondence with $\widetilde r_0$. The values of $\sigma$-meson mass in
Table \ref{tab:table1} look quite reasonable as well. However, the corresponding quantities are
strongly different at small quark numbers in the droplet. We know from the experiments that in the
nuclear matter some increase of the nuclear density is observed. It becomes quite noticeable for
the Helium and is much larger than the standard nuclear density for the Hydrogen.

Obviously, we understand the Thomas--Fermi approximation which is used for
estimating becomes hardly justified at small number of quarks, and we should deal with the
solutions of complete equation system (\ref{sys}). However, one very encouraging hint comes from
the chiral soliton model of nucleon \cite{BB}, where it has been demonstrated that solving this
system (\ref{sys}) the good description of nucleon and $\Delta$ can be obtained.
%%%%%%%%%%%%%%%%%%%%%%%%%%%%%%%%%%%
%%%%%%%%%%%%%%%%%%%%%%%%%%%%%%%%%%%%%%%%%%%%%
%\begin{figure*}[!tbh]
\begin{figure}[!tbh]
\begin{center}
\leavevmode
%\vspace{-4cm}
\includegraphics[width=0.4\textwidth]{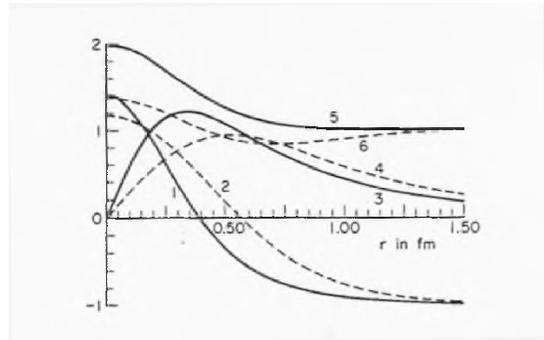}
\caption{The $\sigma$- and $\pi$- fields in units of $F_\pi$.
Curves 1 and 2 show $\sigma$, 3 and 4 show $\pi$, 5 and 6 show
$\sigma^2 +\pi^2$, (curves 1 and 3  with vector and axial-vector fields
contributions included, see Ref. \cite{BB}).}
%ad2
\label{fad2}
\end{center}
\end{figure}
%\end{figure*}
%%%%%%%%%%%%%%%%%%%%%%%%%%%%%%%%%%%%%%%%%%%%%
In a sense we consider an alysing of just three quark system as a
central result of our paper. Looking at Fig. 1 of Ref. \cite{BB}
(we represent it here as Fig. \ref{fad2}) we see that
the curve describing behavior of scalar field at large distances reaches its
minimal value (according to the sign choice done in \cite{BB} it corresponds
to the largest quark mass of order $300$ MeV). It looks like by appoaching
 the center of barion a chiral symmetry is partially restored and a
scalar field in the region of $\sim 0.5$ fm disappears. One of the possible
scenario for solving the system equations (\ref{sys}) could be a variant
in which a scalar field reaches maximal (zero) value  (with a zero value of derivative
over coordinat) at this (or center of baryon) point. Then a scalar field can, in principle,
smoothly approach to its minimal value coming to center of baryon. It allows us conclude
that we could deal with an "ordinary"\ quark with positive (zero) mass for
the solutions of such a type. However, baryon is getting to large widht (size)
in this scenario. There is another type of solutions, in which a "speed"\
of passing by the point $0.5$ fm is not getting slover. In fact such a situation
could realized by doing a chiral rotation where a quark inside a baryon
falls in the metastable region of negative quark masses. Such
a solution develops already quite suitable width of order $\sim 1$ fm
due to presence of massive ($1$ GeV) scalar field. Clearly, the problem
of existance of so heavy  $\sigma$-meson (strengthening the chiral effect)
is crucial to collect a necessary information
on a phase diagram of strongly interacting matter.
 Such solutions develop the surface tension coefficient
which is larger in factor two than the corresponding coefficient of single kink and as we
believe signal some instability of a single kink solution.
%%%%%%%%%%%%%%%%%%%%%%%%%%%%%%%%%%%%%%%%%%%%%%%%%%%%

%%%%%%%%%%%%%%%%%%%%%%%%%%%%%%%%%%%%%%%%%%%%%
%\begin{figure*}[!tbh]
\begin{figure}[!tbh]
\begin{center}
\leavevmode
%\vspace{-4cm}
\includegraphics[width=0.3\textwidth]{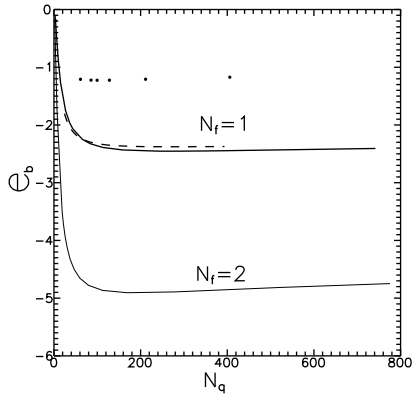}
\caption{The specific binding energy at $N_f=1$ and $N_f=2$ in MeV as a
function of quark number $N_q$.}
%21
\label{f8i}
\end{center}
\end{figure}
%\end{figure*}
%%%%%%%%%%%%%%%%%%%%%%%%%%%%%%%%%%%%%%%%%%%%%
%III_9
Fig. \ref{f8i} displays the specific binding energy of ensemble. It is
defined by the expression similar to Eq. (\ref{sur}) in that the integration over the quark droplet
volume is performed. The specific energy is normalized (compared) to the ensemble energy at the
spatial infinity, i.e. in vacuum. Actually, Fig. \ref{f8i} shows several curves in the upper part of
plot which correspond the calculations with $N_f=1$. The solid line is obtained by
scanning over parameter $\eta$ and corresponds to the data presented in Table \ref{tab:table1}. The
dashed curve is calculated at fixed $\eta=0.4$ but by scanning over parameter
$\stackrel{*}M$. It is clearly seen if the specific energy data are presented as a function of quark number $N_q$
then the solutions, which we are interested in, rally in the local vicinity of the curve where
the maximal binding energy -- $|{\cal E}_b|$ is reached.
The similar solution scanning can be performed over the central density
parameter $\rho_0$ in origin. The corresponding data are dotted for a certain fixed $\stackrel{*}M$
and $\rho_0$. It is interesting to notice that at scanning over any variable discussed a
saturation property is observed and it looks like the minimum in $e_b$ at $N_q\sim 200$--$250$. The
results for the specific binding energy as a function of particle number are in the
qualitative agreement with the corresponding experimental data. And one may say even about the
quantitative agreement if the factor $3$ (the energy necessary to remove one baryon) is taken into account.
 In fact, the equation system (\ref{sys}) represents an equation of balance for
the current quarks circulating between liquid and gas phases.

%III_11
Summarizing we would like to emphasize that in the present paper we have
demonstrated how a phase transition of liquid--gas kind (with the reasonable values of
parameters) emerges in the NJL-type models. The constructed quark ensemble displays some interesting
features for the nuclear ground state (for example, an existence of the state degenerate with
the vacuum one), and the results of our study are suggestive to speculate that the quark
droplets could coexist in equilibrium with vacuum under the normal conditions. These droplets
manifest themselves as bearing a strong resemblance to the nuclear matter.

\section{Conclusion}
In the present paper we described quantum liquids (Landau Fermi-liquids)
resulting from the quark models with four-fermion interaction. This consideration is based on
the identity of results obtained in \cite{MZ2} by using a dressing Bogolyubov transformation
and mean field approximation. We demonstrated that the mean energy of ensemble serves as an
energy functional of the Landau theory. It was shown that in a wide range of potentials
interesting for applications one can expect the quantum liquids to behave in the essentially
same way. For some of their properties a band of estimates was obtained. A comparison of NJL and
KKB models, substantially different in many aspects, demonstrates that the properties of
quantum liquids do not actually depend on a shape of the formfactor (a natural interaction
length); rather, they are mainly determined by the coupling constant of interaction. It was shown
that a common distinctive feature of ensembles is a presence of occupied states degenerate
with respect to vacuum in chemical potential and pressure. Taking this observation the
inhomogeneous states, which allowed at describing a transition layer, estimating a surface tension,
as well as studying some properties of quark liquid droplets, were considered. It is
noted that in the case of a small number of quarks in a droplet instability associated with lowering
of the energy barrier, separating chiral phases, apparently manifests itself. This
instability is seen in two kinks merging into one chiral soliton. An idea of dynamical equilibrium of a
mixed phase consisting of baryon matter and vacuum was discussed as a possible scenario
for explaining stability of nuclear matter.

ACKNOWLEDGMENTS

Authors are indebted D.V. Anchishkin, K. A. Bugaev, E.-M. Ilgenfritz, V.V.
Skalozub, A. M. Snigirev and many other colleagues for numerous fruitful discussions.
S.V.M. is deeply grateful to Professor L.V. Keldysh for interesting discussion. The work is
supported by the Special Project of the Physics and Astronomy Division of the National Academy
of Sciences of Ukraine.

%\newpage

\end{document}